\UseRawInputEncoding
\documentclass[
reprint,
superscriptaddress,
amsmath,amssymb,
aps,
prl,
]{revtex4-2}
\usepackage{graphicx}
\usepackage{dcolumn}
\usepackage{mathrsfs}
\usepackage{bm}
\begin{document}
\title{Electric Quadrupolar Contributions
in the Magnetic Phases of UNi$_4$B
}
\author{T. Yanagisawa}
 \affiliation{Department of Physics, Hokkaido University, Sapporo 060-0810, Japan}
\author{H. Matsumori}
 \affiliation{Department of Physics, Hokkaido University, Sapporo 060-0810, Japan}
\author{H. Saito}
 \affiliation{Department of Physics, Hokkaido University, Sapporo 060-0810, Japan}
\author{H. Hidaka}
 \affiliation{Department of Physics, Hokkaido University, Sapporo 060-0810, Japan}
\author{H. Amitsuka}
 \affiliation{Department of Physics, Hokkaido University, Sapporo 060-0810, Japan}
\author{S. Nakamura}
 \affiliation{Institute for Materials Research, Tohoku University, Sendai 980-8577, Japan}
\author{S. Awaji}
 \affiliation{Institute for Materials Research, Tohoku University, Sendai 980-8577, Japan}
\author{D. I. Gorbunov}
 \affiliation{Hochfeld-Magnetlabor Dresden (HLD-EMFL) and W\"urzburg-Dresden Cluster of Excellence ct.qmat, Helmholtz-Zentrum Dresden-Rossendorf (HZDR), 01328 Dresden, Germany}
\author{S. Zherlitsyn}
 \affiliation{Hochfeld-Magnetlabor Dresden (HLD-EMFL) and W\"urzburg-Dresden Cluster of Excellence ct.qmat, Helmholtz-Zentrum Dresden-Rossendorf (HZDR), 01328 Dresden, Germany}
\author{J. Wosnitza}
 \affiliation{Hochfeld-Magnetlabor Dresden (HLD-EMFL) and W\"urzburg-Dresden Cluster of Excellence ct.qmat, Helmholtz-Zentrum Dresden-Rossendorf (HZDR), 01328 Dresden, Germany}
 \affiliation{Institut f\"{u}r Festk\"{o}rper- und Materialphysik, TU Dresden, 01062 Dresden, Germany}
\author{K. Uhl\'{i}\v{r}ov\'{a}}
 \affiliation{Department of Condensed Matter Physics, Faculty of Mathematics and Physics, Charles University, 121 16 Prague 2, Czech Republic}
 \author{M. Vali\v{s}ka}
 \affiliation{Department of Condensed Matter Physics, Faculty of Mathematics and Physics, Charles University, 121 16 Prague 2, Czech Republic}
 \author{V. Sechovsk\'{y}}
 \affiliation{Department of Condensed Matter Physics, Faculty of Mathematics and Physics, Charles University, 121 16 Prague 2, Czech Republic}
\date{\today}

\begin{abstract}
We present acoustic signatures of the electric quadrupolar degrees of freedom in the honeycomb-layer compound UNi$_4$B. The transverse ultrasonic mode $C_{66}$ shows softening below 30 K both in the paramagnetic phase and antiferromagnetic phases down to $\sim0.33$ K. Furthermore, we traced magnetic field-temperature phase diagrams up to 30 T and observed a highly anisotropic elastic response within the honeycomb layer. These observations strongly suggest that $\Gamma_6$(E$_{\rm 2g}$) electric quadrupolar degrees of freedom in localized $5f^2$ ($J = 4$) states are playing an important role in the magnetic toroidal dipole order and magnetic-field-induced phases of UNi$_4$B, and evidence some of the U ions remain in the paramagnetic state even if the system undergoes magnetic toroidal ordering.
\end{abstract}
\maketitle

The multipole formulation and its foundational concept in solid-state physics have been developed by intensive research on $f$-electron systems~\cite{Santini2009, Kuramoto2009}. Recently, new theories based on the common language of `multipoles'~\cite{Kusunose2008,Kuramoto2008} and `augmented multipoles'~\cite{HayamiPRB2020,HayamiPRB2018,HayamiJPSJ2018, Watanabe2018, Suzuki2018}, which are spatially extended multipoles, have been evoked to construct a new framework for understanding various physical phenomena that are related to spin-orbit interactions beyond the differences in electron orbitals. In particular, the odd-parity augmented multipoles, including magnetic/electric and toroidal ones~\cite{Yatsushiro2020, Yatsushiro2019, Hayami2014}, have recently been extensively studied. Recent academic advances in understanding augmented multipoles have been preceded by theory rather than experiment. Therefore, it is necessary to demonstrate whether the new framework allows for a unified understanding of spontaneous spatial inversion symmetry breaking in metallic and insulating compounds. A major experimental challenge is to demonstrate odd-parity multipole ordering by observing cross-correlation phenomena and spontaneous spatial inversion-symmetry breaking in a suitable compound ~\cite{HayamiPRB2020,Watanabe2018}. Among them we focus on the U-based honeycomb-layer compound UNi$_4$B~\cite{Mydosh1992,Mentink1993,Mentink1994,Mentink1995}, which is considered to be a good candidate for studying augmented odd-parity multipoles, magnetic toroidal dipoles (MTDs), and the interplay with magnetoelectric phenomena~\cite{Hayami2014, Saito2018}.

\begin{figure}[b]
\includegraphics[width=1.0\linewidth]{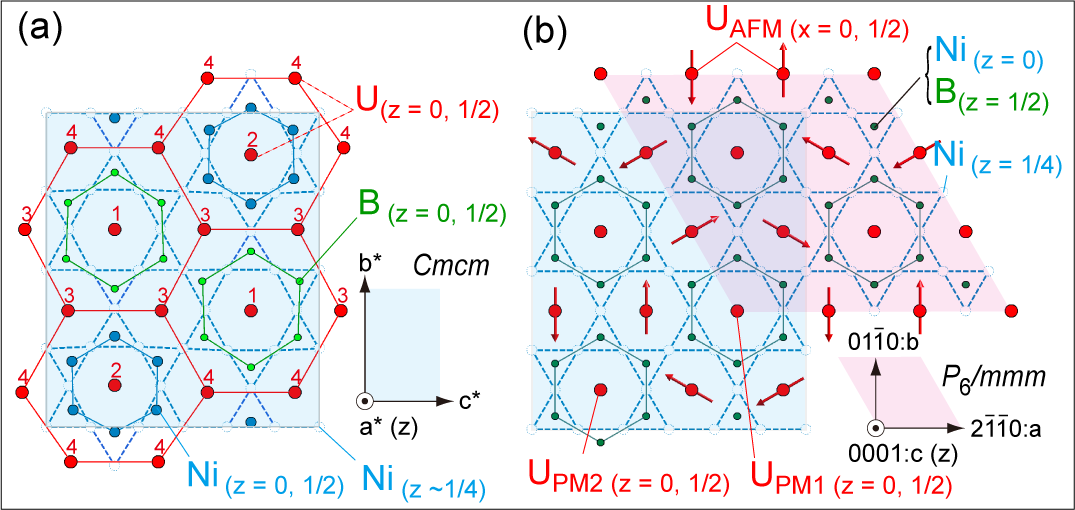}
\caption{\label{fig:fig1}  Crystal and magnetic structure of UNi$_4$B, reported for $B$ = 0~\cite{Haga2008, Mentink1995}. (a) The pseudo-honeycomb network consists of 2/3 of U ions with $Cmcm$ lattice. Red, blue, and green circles indicate U and Ni or B on the layer at $z$ = 0, 1/2, respectively. Dashed lines with open circles show the (pseudo) Kagome layer of Ni atoms at $z \sim$ 1/4. (b) Colored backgrounds with pink-rhomboidal and blue-rectanglar shapes denote antiferromagnetic unit cells in hexagonal ($P6/mmm$) and orthorhombic ($Cmcm$) symmetry, respectively.}
\end{figure}
UNi$_4$B crystallizes in an orthorhombic structure (Space group; $Cmcm$, $D_{2h}^{17}$, No. 63) as shown in Fig. 1(a)~\cite{Haga2008}. Below $T_{\rm N}$ = 20.4 K, this compound orders antiferromagnetically (AFM) in a magnetic structure where the magnetic moments are carried by two thirds of the U ions [U$_{\rm AFM}$ sites in Fig. 1(b) form vortices in each pseudo-honeycomb plane], and one third of the U ions [U$_{\rm PM1}$ or U$_{\rm PM2}$ in Fig. 1(b)] remain in a paramagnetic (PM) state.~\cite{Mydosh1992, Mentink1994, Mentink1995} Assuming a hexagonal crystal structure ($P6/mmm$, $D_{6h}^1$, No. 191), an exotic magnetic structure was proposed from neutron scattering experiments in earlier studies~\cite{Mentink1995}. Since a slight deformation of the crystal structure from hexagonal to orthorhombic symmetry and different site occupations of Ni and B atoms have recently been reconfirmed by neutron and resonant X-ray scattering studies as well as by $^{11}$B-NMR measurements~\cite{Tabata, Takeuchi2020}, the previously proposed magnetic structure should be reconsidered based on the orthorhombic space group.

On the other hand, Hayami {\it et al.} has pointed out that such vortex-type magnetic structure in the (pseudo) honeycomb arrangement in UNi$_4$B can be understood in the framework of MTD order (see Fig. S1 in the supplemental material (SM))~\cite{Hayami2014, SM}. Their theory has also predicted a new magnetoelectric effect: current-induced magnetization, which can occur in ferro-toroidal ordered metallic compounds, which has been experimentally confirmed in UNi$_4$B~\cite{Saito2018}. Recently, Yatsushiro and Hayami have reported on a theoretical investigation of an atomic scale MTD by taking into account orbital degrees of freedom with different parity~\cite{Yatsushiro2019}. This theory predicts that the orbital degrees of freedom in an interorbital space play an important role in stabilizing MTD order by odd-parity hybridization. However, the theory deals with the tetragonal point group $C_{\rm 4v}$, and the contribution of the orbital degrees of freedom such as even-parity electric multipolar moments to the noncollinear magnetic order in UNi$_4$B has not been investigated.

Another fascinating point and also an open question for this compound is a specific-heat anomaly at $T^*$ $\sim0.33$ K of unknown origin~\cite{Movshovich1999}. Previous studies have explained that the narrowness of the 0.33 K anomaly in UNi$_4$B may be an indication of glassy behavior caused by the geometrical frustration of the paramagnetic U spins and their Kondo screening by the conduction electrons, since AC magnetic susceptibility and $\mu$SR measurements have shown no changes in the magnetic structure below $T^*$~\cite{Movshovich1999}. Other possibilities, which have not been verified yet, are a nonmagnetic multipolar order of the PM-1/3 U ions and/or Schottky peak due to level splitting of the degenerate CEF ground state with low orthorhombicity. In order to study the electrical multipole contribution, it is useful to measure the elastic constants by ultrasound~\cite{Luethi}. The elastic constants reflect the coupling of the strain field caused by the ultrasonic wave to the electric multipolar moments, which are described by orbital degrees of freedom of the CEF state. In this study, we show evidence of an electric multipolar ground state for UNi$_4$B based on ultrasound results. We further analyzed the possible contributions of electric quadrupoles to the noncollinear magnetic order and to the low-temperature specific-heat anomaly (See Sect. C to G in the SM~\cite{SM} for experimental and analysis details).
\begin{figure}
\includegraphics[width=1.0\linewidth]{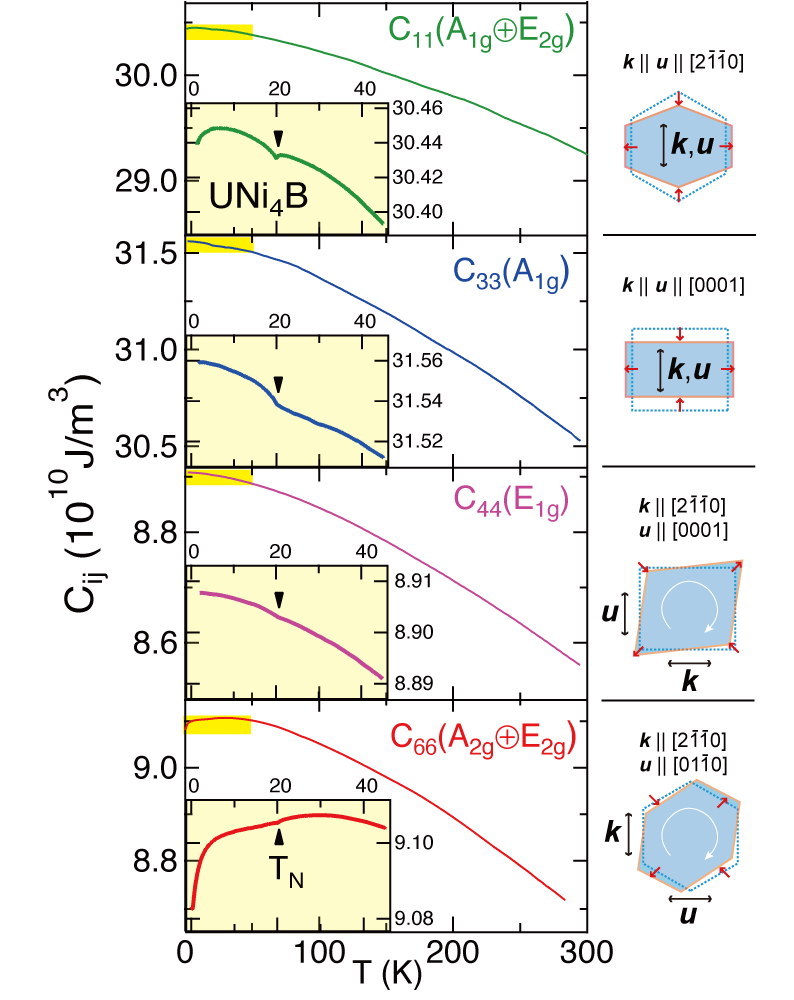}
\caption{\label{fig:fig2} Elastic constants of UNi$_4$B as a function of temperature. The insets in each panel show enlarged views of the data below 50 K. Illustrations of the distorted hexagon and/or rectangle indicate the lattice strain, which is induced by the respective ultrasonic mode (See Table SI in the SM) ~\cite{SM,Fulde, Goto}}
\end{figure}
\begin{figure}
\includegraphics[width=0.9\linewidth]{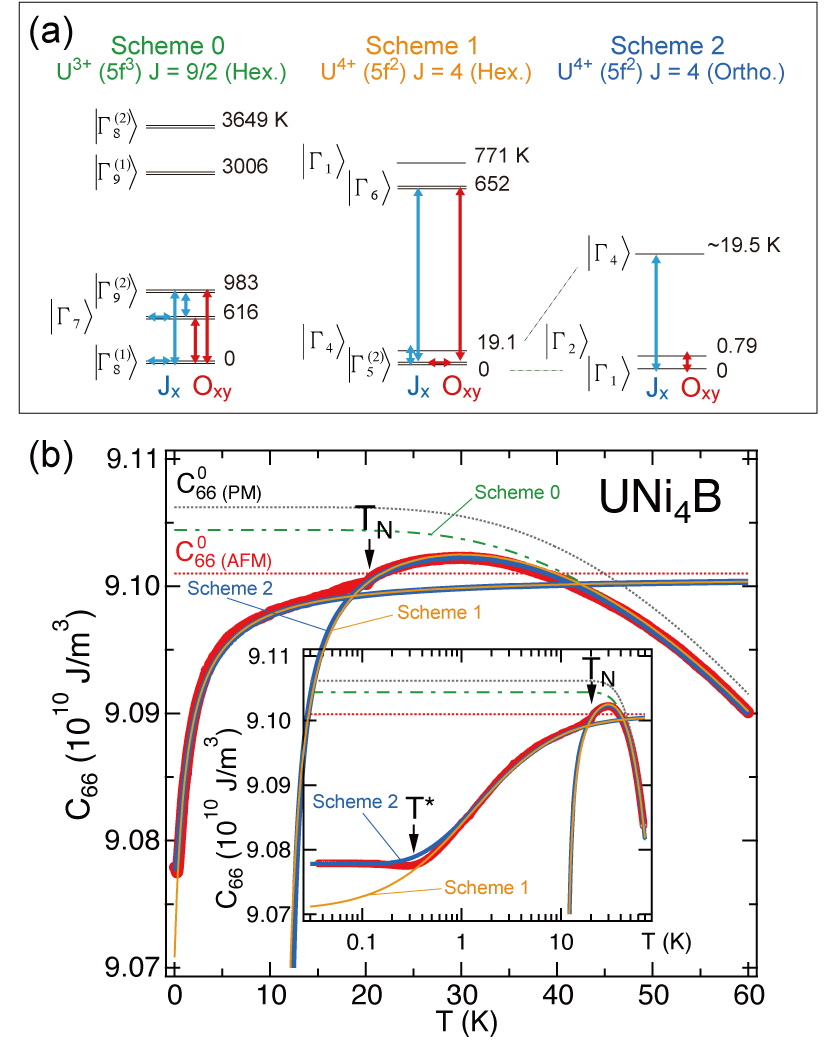}
\caption{\label{fig:fig3} (a) CEF level schemes used for the present analysis (see Table SIII, SIV, and SV in the SM)~\cite{SM}. Blue and red arrows indicate selection rules where finite matrix elements exist for magnetic dipole $J_x$ and electric quadrupole $O_{xy}$, respectively. (b) Temperature dependence of the elastic constant $C_{66}$, compared with calculations based on the quadrupolar susceptibility using the CEF level schemes shown in panel (a): Scheme 0 (green), Scheme 1 (orange), and Scheme 2 (blue) as shown by dashed and solid curves. The dotted black and red curves show the background of the elastic constant $C_{66}^0$ in the PM and AFM phases, respectively. The inset shows an enlarged view of the low-temperature region on a log-$T$ scale.}
\end{figure}

Figure 2 shows the measured elastic constants of UNi$_4$B as a function of temperature. Here, the four ultrasonic modes are symmetrized using the hexagonal point group $D_{6h}$. The ultrasound induces local strain and rotation fields~\cite{SM} in the solid sample as shown by the schematic illustrations in each panel of Fig. 2. The local strain and rotation field behave as conjugate fields for electric quadrupole or electric hexadecapole moments ~\cite{Kurihara2017}. Responses of the multipoles can be observed as sound-velocity change and ultrasonic attenuation via electron-phonon interaction. 
By comparing the temperature dependence of the four ultrasonic modes, it becomes obvious that only the transverse ultrasonic mode $C_{66}$ exhibits a softening below 30 K in the PM phase with a kink at $T_{\rm N}$ and keeps decreasing in the antiferromagnetic (AFM) phase down to $\sim0.33$ K [Fig. 3(b)].
\begin{figure*} [t]
\includegraphics[width=1.0\linewidth]{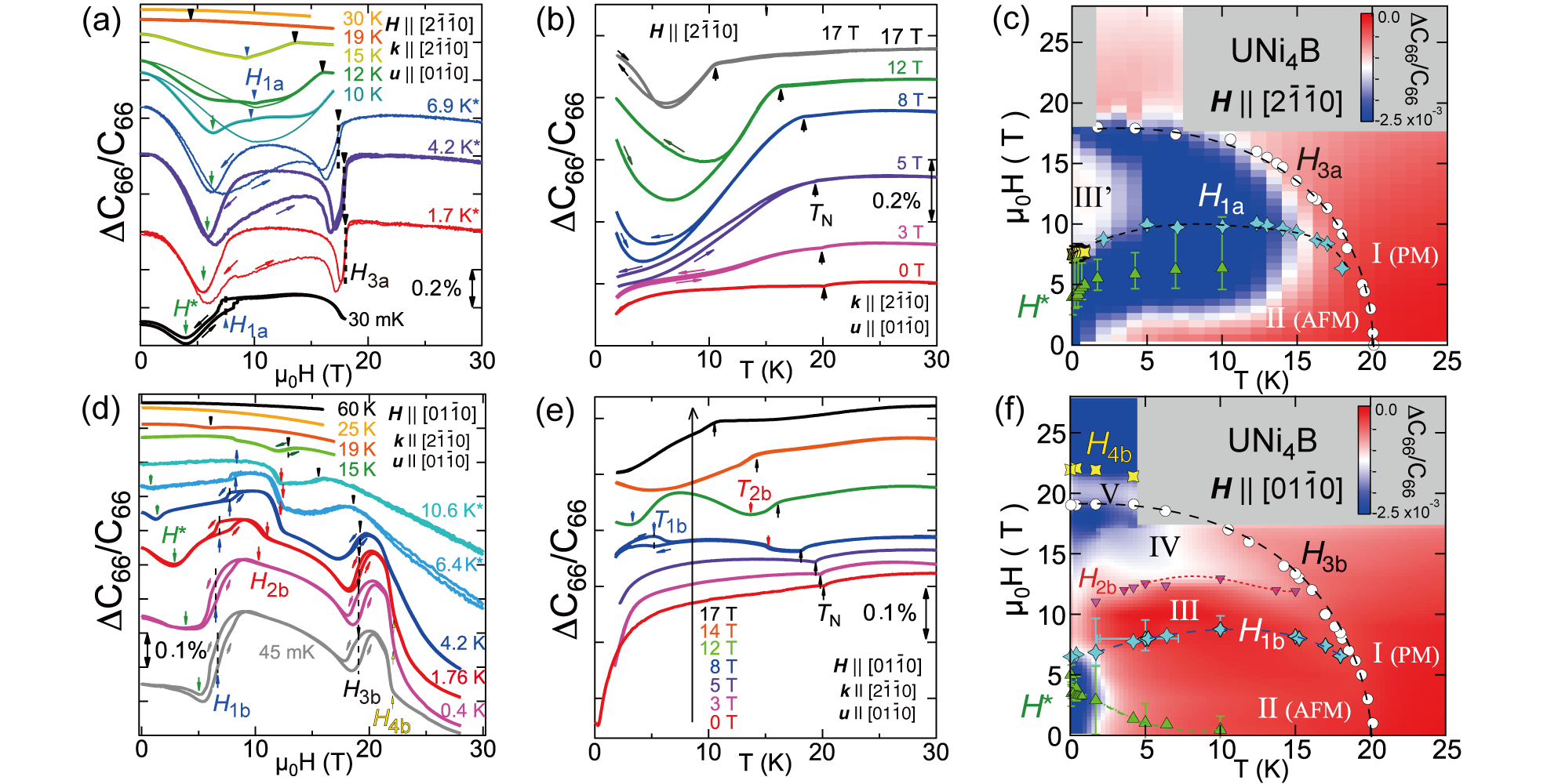}
\caption{\label{fig:fig4} Magnetic-field dependence of $C_{66}$ for (a) $H \parallel$ [$2\bar{1}\bar{1}0$] and (d) $H \parallel$ [$01\bar{1}0$] of UNi$_4$B at selected temperatures. $C_{66}$ vs. temperature under various magnetic fields for (b) $H \parallel$ [$2\bar{1}\bar{1}0$] and (e) $H \parallel$ [$01\bar{1}0$]. In each panel, vertical arrows indicate elastic anomalies, which correspond to phase boundaries in the magnetic field-temperature phase diagram of UNi$_4$B for (c) $H \parallel$ [$2\bar{1}\bar{1}0$] and (f) $H \parallel$ [$01\bar{1}0$]. Dotted and dashed lines are guides for the eyes. The relative change of the elastic constant $C_{66}$ is highlighted by the color map from zero (red) to large negative (blue) values.}
\end{figure*}

The other ultrasonic modes, the longitudinal $C_{11}$, $C_{33}$, and transverse $C_{44}$ modes, do not show such softening particularly below $T_{\rm N}$. From the selection rules within the category of even-parity multipoles, the results indicate that an electric quadrupole with $\Gamma_6$(E$_{\rm 2g}$) symmetry (in the hexagonal point group) is active in UNi$_4$B. In the first stage of the analysis, we consider the conventional localized $5f$-electronic states with even-parity CEF levels and multipoles for the analysis of the elastic responses. Here, we do not take into account the contribution from odd-parity multipoles, because such effect only couples to the elastic strain through cross-correlation with the application of appropriate external fields with odd-parity space-inversion symmetry~\cite{HayamiPRB2018, Watanabe2018}. As shown below, the current analysis mimics the experimental results well that the odd-parity mixing inherent in the CEF is negligible.
The softening in $C_{66}$ in the paramagnetic phase could, however, not be reproduced by the previously proposed CEF level scheme [Scheme 0 in Fig. 3(a)] ~\cite{Oyamada2009} with the localized $5f$-electronic state of U$^{3+}$ having a hexagonal symmetry because the $5f^3$ ($J = 9/2$) state only shows $\Gamma_6$-quadrupolar excitation ($O_{xy}=\sqrt{3}(J_{x}J_{y}+J_{y}J_{x})/2$) in the off-diagonal elements between the ground-state Kramers doublet and the excited levels that are separated by an energy gap of over 600 K, as shown in Fig. 3(a) (for matrix elements, see Sect. F in the SM~\cite{SM}).

In order to reproduce the softening in the $C_{66}$ mode at higher temperature than $T_{\rm N}$, we propose a different CEF model [Scheme 1 in Fig. 3(a)] with a localized $5f^2$ ($J = 4$) state of U$^{4+}$, which has a pseudo-triplet ground state. The CEF parameters of CEF Scheme 1 (Table SIV in the SM~\cite{SM}) are set to reproduce simultaneously the elastic softening in $C_{66}$, no elastic softening in $C_{44}$, and also the magnetic susceptibility below 50 K at the same time (for details see Fig. S3 in the SM~\cite{SM}). The coupling constant $g'_{\Gamma}$ of quadrupolar inter-site interactions is described by the Hamiltonian
$\mathcal{H}_{\rm MM}=-\sum_{\alpha}g'_{\Gamma_6}\left<O_{xy}\right>O_{xy}^{\alpha}$ for sublattices $\alpha$ and a $\Gamma_6$ symmetry quadrupolar moment $O_{xy}$ (see Eqs. 17-21 in the SM) \cite{Thalmeier_Luthi91, Luethi, SM}. Our analysis reveals a positive value $g'_{\Gamma 6{\rm(PM)}}$ = +0.42 K, which strongly suggests the presence of a weak but finite ferro-type quadrupolar interaction in the PM phase of UNi$_4$B. A possible relationship between the ferro-type MTD order of this system and the ferro-quadrupolar interaction is nontrivial and remains an open question. The softening of $C_{66}$ in the AFM phase can also be analyzed using the same CEF parameters, since the PM-1/3 U ions (on the U$_{\rm PM1}$ or U$_{\rm PM2}$ sites) located in the center of the pseudo-honeycomb plane are not affected by the on-site electric/magnetic fields formed by the MTD moment. Although the global inversion symmetry on the U$_{\rm PM1}$ and U$_{\rm PM2}$ sites is broken due to the MTD order, the odd-parity CEF states and multipole will, however, not be active on the PM-1/3 U sites when only considering the $J = 4$ Hilbert space and assuming weak orbital coupling between the U and Ni or B ions.

The inset in Fig. 3(b) shows the temperature dependence of $C_{66}$ on a log-$T$ scale. Calculations of the quadrupolar susceptibility of the PM-1/3 U ions using the CEF Scheme 1 (orange curve)~\cite{SM} with hexagonal symmetry well reproduces the softening in the AFM phase down to $\sim0.33$ K. Here, the contribution from the AFM ordered-2/3 U ions is assumed as constant background and the CEF level scheme of the PM-1/3 U ions is not changed from PM phase. Note that the intersite quadrupolar interaction for the AFM phase obtained by our analysis is negative with $g'_{\Gamma 6{\rm(AFM)}}=-0.045$ K, which means the presence of antiferro-quadrupolar (AFQ) interaction in the AFM phase. In order to reproduce the temperature dependence down to $T^*$ in the present analysis, it is unlikely that $g'_{\Gamma 6{\rm(AFM)}}$ takes positive value. Therefore, we can conclude that in the ordered phase there is an antiferro-quadrupolar interaction between the U$_{\rm PM}$'s [in Fig. 1(b)], which is fundamentally the opposite to that for U$_{\rm AFM}$ in the PM phase.

The calculated result using Scheme 1 deviates from the experimental data below $T^*\sim0.33$ K. This deviation indicates a small CEF splitting of the ground-state doublet. Such splitting might occur for two reasons: 1) symmetry lowering due to ordering of the remaining PM-1/3 U multipolar moments or 2) the crystal structure essentially having lower symmetry. Since it has been confirmed that the point group of the U ions in the present system has orthorhombic symmetry, we have modified the CEF Scheme 1 by adding low orthorhombicity (as CEF parameter $B_2^2$ and $B_4^2$ for Steven's operator $O_2^2$ and $O_4^2$ (see Eqs. 3-11 in the SM)~\cite{Stevens_1952, HUTCHINGS1964, Kusunose2008, SM}, respectively, have finite values) to split the ground-state non-Kramers doublet with a gap of $\it{\Delta}\sim$ 0.79 K. This CEF Scheme 2 [Fig. 3(a), Table SV in the SM] reproduces the leveling off of $C_{66}$ (blue curve) as well as a Schottky-type specific-heat peak at $\it{\Delta}/$ 2.398 $\sim$ 0.33 K~\cite{Movshovich1999}. We, therefore, conclude that the mentioned $T^*$ transition, found in specific heat, is a Schottky anomaly due to the orthorhombicity of the crystal. It should be noted that we determined the magnetic-field dependence of $T^*$ and its anisotropy for $H \parallel$ [$2\bar{1}\bar{1}0$] and [$01\bar{1}0$] in the elastic constant $C_{66}$. Our results are roughly consistent with those of earlier studies~\cite{Mentink1995} (see Fig. S8 in the SM~\cite{SM}). 

The temperature and magnetic-field dependence of the elastic constant $C_{66}$ are shown in Figs. 4(a) and 4(b) for $H \parallel$ [$2\bar{1}\bar{1}0$] and in Figs. 4(d) and 4(e) for $H \parallel$ [$01\bar{1}0$]. In Fig. 4(a), the data with an asterisk are obtained in pulsed-magnetic-field measurements up to 60 T (see Fig. S9 in the SM~\cite{SM}). We obtained $C_{66}$ vs. $H \parallel$ [$01\bar{1}0$] up to 28 T, shown in Fig. 4(d), in static magnetic fields using the cryogen-free hybrid magnet system equipped with a dilution refrigerator \cite{IMR}. Several elastic anomalies are observed, which are indicated by arrows; $H^*$, $H_{\rm 1b}$, and $H_{\rm 3b}$ for $H \parallel$ [$2\bar{1}\bar{1}0$], $H^*$, $H_{\rm 1a}$, $H_{\rm 2a}$, $H_{\rm 3a}$, and $H_{\rm 4a}$ for $H \parallel$ [$01\bar{1}0$]. The data display both up and down sweeps of the magnetic field. We observe hysteretic regions below $H_{\rm 3a}$. The elastic responses in $C_{66}$ show a large in-plane (0001) anisotropy for $H \parallel$ [$2\bar{1}\bar{1}0$] and $H \parallel$ [$01\bar{1}0$], while the magnetization does not show such strong anisotropies~\cite{Mentink1995}. The positions of the elastic anomalies are indicated as well in the magnetic field-temperature ($H$-$T$) phase diagrams as shown in Figs. 4(c) and 4(f). Here, a number of phases are distinguishable for $H \parallel$ [$01\bar{1}0$]; PM phase I, AFM phase II, the spin-reoriented AFM phase III, the spin-flop phase IV with hysteresis, which was previously evidenced by magnetization data~\cite{Mentink1995}, and a newly found unknown phase V. The obtained phase boundaries are consistent with the previously reported phase diagram ~\cite{Mentink1995} except for the high-magnetic-field region. On the other hand, the $H$-$T$ phase diagram for $H \parallel$ [$2\bar{1}\bar{1}0$] is quite different with less phases and completely different elastic responses. The background red-white-blue color code in Figs. 4(c) and 4(f) represents the relative changes in $C_{66}$ from larger to smaller stiffness. Remarkably, the contour plot shows a significant difference in stiffness for $H \parallel$ [$01\bar{1}0$] and [$2\bar{1}\bar{1}0$], though no difference was detected in magnetization. This new observation clearly indicates a possible contribution of electric quadrupoles of the PM-1/3 U ions, which modifies the spin-reorientation process as well.

In the low-magnetic-field and low-temperature regions for both $H \parallel$ [$01\bar{1}0$] and $H \parallel$ [$2\bar{1}\bar{1}0$], the blue color indicates enhanced contributions from the $\Gamma_6$(E$_{\rm 2g}$)-electric quadrupoles, i.e., this system has incoherent fluctuations of the electric quadrupoles due to the CEF pseudo-doublet ground state. In general, the ground-state doublet (including non-Kramers doublet) splits due to the Zeeman effect (with mixing of excited-levels wave functions) in an external magnetic field or in the internal fields produced by magnetic order, and the quadrupole degrees of freedom become inactive, resulting in a hardening of $C_{66}$. Remarkably, $C_{66}$ experiences a softening in phase IV for $H \parallel$ [$01\bar{1}0$] and in the intermediate temperature range of phase III' for $H \parallel$ [$2\bar{1}\bar{1}0$] compared with the changes in the other phases. This fact suggests a reactivation of the quadrupole degrees of freedom with $\Gamma_6$(E$_{\rm 2g}$) symmetry above $\sim$12 T in the low-temperature region for $H \parallel$ [$01\bar{1}0$].

In summary, we conclude that the electric-quadrupole degrees of freedom play a crucial role in the low-temperature properties of UNi$_4$B, leading to anisotropic $H$-$T$ phase diagrams and a newly revealed field-induced phase V. The observed softening of the $C_{66}$ elastic constant can be well explained by quadrupolar-strain interactions. The corresponding CEF analysis results in a new level scheme (Scheme 2) taking into account the established orthorhombic symmetry and the $5f^2$ ($J = 4$) state of uranium ions. Furthermore, in this level scheme the puzzling specific-heat anomaly at $\sim0.33$ K can be understood as a Schottky anomaly due to a small level splitting of the non-Kramers ground-state doublet. Moreover, our results confirm that some of the U ions stay disordered in the MTD ordered phase. Further theoretical considerations would be of interest to clarify the quadrupolar contributions in the toroidal order and anisotropic elastic response in the newly established magnetic-field-temperature phases.
\begin{acknowledgments}
We thank C. Tabata, S. Hayami and H. Kusunose for fruitful discussions. The present research was supported by JSPS KAKENHI Grant Nos. JP17K05525, JP15KK0146, JP15K13509, JP15H05885, JP15K21732, and the Strategic Young Researcher Overseas Visits Program for Accelerating Brain Circulation from JSPS. This study was partly supported by Hokkaido University, Global Facility Center (GFC), Advanced Physical Property Open Unit (APPOU), funded by MEXT under “Support Program for Implementation of New Equipment Sharing System” Grant No. JPMXS0420100318. A part of this work was performed at the High Field Laboratory for Superconducting Materials, Institute for Materials Research, Tohoku University (Project No. 17H0063). We also acknowledge the support of the Hochfeld-Magnetlabor Dresden at HZDR, a member of the European Magnetic Field Laboratory (EMFL), and the Deutsche Forschungsgemeinschaft (DFG) through SFB 1143 (Project No.\ 247310070) and the W\"{u}rzburg-Dresden Cluster of Excellence on Complexity and Topology in Quantum Matter--$ct.qmat$ (EXC 2147, Project No.\ 390858490). The sample used in HZDR has been grown and characterized in the Materials Growth and Measurement Laboratory MGML which is supported  within  the  program  of  Czech  Research Infrastructures (Project No. LM2018096).
\end{acknowledgments}

\renewcommand{\thesection}{\Roman{section}}.
\renewcommand{\thetable}{S\Roman{table}}
\renewcommand{\thefigure}{S\arabic{figure}}
\setcounter{table}{0}
\setcounter{figure}{0}
\onecolumngrid
\begin{center}
\vspace{5mm}
\newpage
{\bf \large Supplemental Material for\\ 
Electric Quadrupolar Contributions\\ 
in the Magnetic Phases of UNi$_4$B}\\
\vspace{5mm}

T. Yanagisawa, H. Matsumori, H. Saito, H. Hidaka, H. Amitsuka, S. Nakamura, S. Awaji, D. I. Gorbunov, S. Zherlitsyn, J. Wosnitza, K. Uhl\'{i}\v{r}ov\'{a}, M. Vali\v{s}ka, V. Sechovsk\'{y}\\
\vspace{5mm}
\end{center}
\twocolumngrid
\section*{\label{sec:Schematic}A.	Schematic Illustration of the Magnetic Toroidal Order}
Fig. S1 shows a schematic illustration of the magnetic toroidal order in UNi$_4$B with hexagonal structure. Green and red arrows indicate the magnetic dipolar moment and the magnetic toroidal moment, respectively. The spheres indicate the U ions, with the blue-colored ions representing the one-third of the U ions that maintain a paramagnetic state in the AFM phase. This vortex-like magnetic structure can be reproduced theoretically by assuming Heisenberg-exchange interactions between spins on the honeycomb arrangement (similar to the 120 degree three-sublattice N\'{e}el order in a triangular lattice) using an appropriate parameter space. Thus, the possible contribution of orbital degrees of freedom on this one-third of U ions is not required in this framework. 

\begin{figure}[h]
\includegraphics[width=1.0\linewidth]{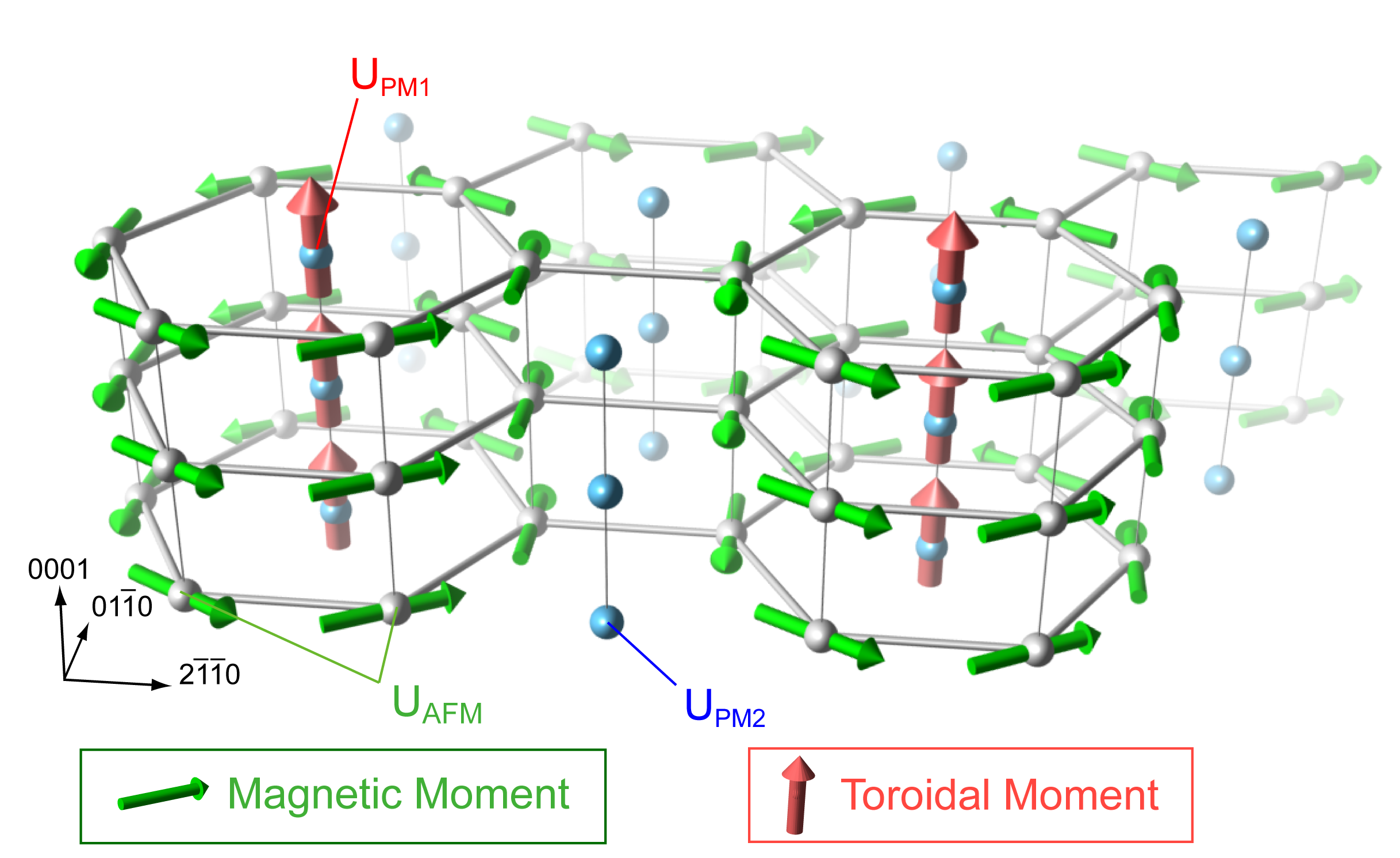}
\caption{\label{fig:figS1}
Scheme of the ferro-magnetic-toroidal order of UNi$_4$B.}
\end{figure}

\section*{\label{sec:Indexing}B.	Indexing A Hexagonal System with Four Axes}
The four axes, ${\bf a_1}$, ${\bf a_2}$, ${\bf a_3}$, and ${\bf c}$, of the hexagonal unit cell are indicated in Fig. S2. When the trace operator ${\bf T}$ is written as ${\bf T} = h{\bf a_1} + k{\bf a_2} + l{\bf a_3} + m{\bf c}$, the high-symmetry axes are described by the four-digit Weber indices as [$hklm$]. Since the symmetry of the axes  ${\bf a_1}$, ${\bf a_2}$, and ${\bf a_3}$ is equivalent and follows the relationship ${\bf a_1}+{\bf a_2}+{\bf a_3}={\bf 0}$, the combination of the indices with relative prime integral numbers is also limited as $h + k +l = 0$.

\begin{figure}[h]
\includegraphics[width=0.8\linewidth]{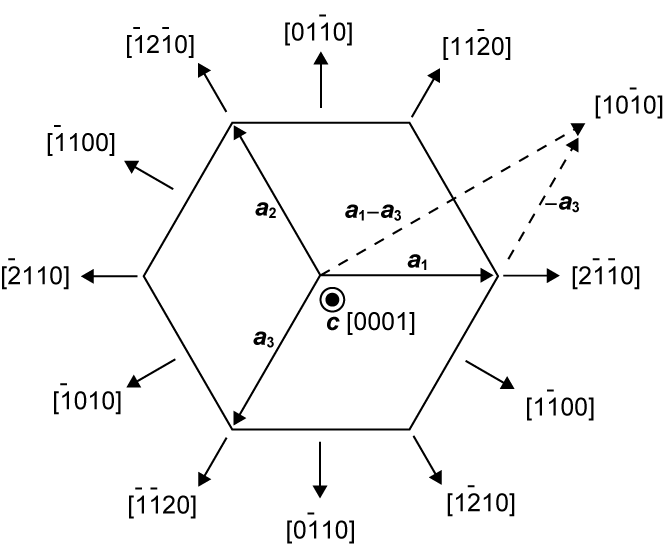}
\caption{\label{fig:figS2}
Direction indices in the hexagonal system with four-digit indices.}
\end{figure}

\section*{\label{sec:Experimental}C.	Experimental Details}
Two single crystals of UNi$_4$B with parallel [$2\bar{1}\bar{1}0$], [$01\bar{1}0$], and [0001] facets were prepared. Here, we used the hexagonal point-group symmetry with a four-axis index for describing lattice directions, since the orthorhombicity of the present crystal is negligibly low ($\sim0.02^\circ$ deviation from $60^\circ$ of the hexagonal lattice) and cannot be distinguished using conventional X-ray diffractometry. Sample 1, which was grown by Czochralski technique, with dimensions of $3.94\times1.96\times1.33$ mm$^3$, was used for measurements under static magnetic fields up to 17 T at Hokkaido University and up to 28 T using the cryogen-free hybrid magnet system at High Field Laboratory for Superconducting Materials, IMR Tohoku University. Sample 2, which was grown by the floating-zone technique at Charles University, with dimensions of $2.23\times1.67\times1.80$ mm$^3$, was used for pulsed-magnetic-field studies up to 60 T with pulse durations of 150 ms at the Dresden High Magnetic Field Laboratory (HLD) at HZDR. No obvious sample dependence was observed at least with X-ray diffraction, magnetization, and ultrasonic measurements. Ultrasound was generated and detected using LiNbO$_3$ transducers with a thickness of 100 $\mu$m, which were bonded on well-polished sample surfaces with RTV silicone or superglue. Sound-velocity measurements were performed by applying a conventional phase-comparative method~\cite{Luethi} using a digital storage oscilloscope. The sound velocities $v_{ij}$(m s$^{-1}$) have been converted to elastic constants $C_{ij}$ (J m$^{-3}$) using the formula $C_{ij} =\rho v_{ij}^2$. Here, $\rho$ = 10.88 (g cm$^{-3}$) is the density of the sample calculated from the lattice constants $a = 4.952$\AA\, and $c = 6.954$\AA.

\section*{\label{sec:Formulation}D.	Formulation of Multipolar Susceptibility}
We start from the Hamiltonian for a hexagonal CEF~\cite{HUTCHINGS1964} and an elastic-strain-mediated perturbed state~\cite{Thalmeier_Luthi91},
\begin{equation}
 \mathcal{H}=\mathcal{H}_{\rm CEF}+\sum_{\Gamma}\frac{\partial \mathcal{H}_{\rm CEF}}{\partial \epsilon_\Gamma}\epsilon_\Gamma.
 \end{equation}
Here, $\epsilon_\Gamma$ is the symmetrized strain with point-group symmetry $\Gamma$, which is induced by ultrasound. The hexagonal CEF Hamiltonian taking Zeeman effect and tiny orthorhombic distortions into consideration is written as
\begin{eqnarray}
 \mathcal{H}_{\rm CEF}&=&B_2^0O_2^0+B_2^2O_2^2+B_4^0O_4^0+B_4^2O_4^2+B_4^4O_4^4\nonumber\\
 &&+B_6^0O_6^0+B_6^2O_6^2+B_6^4O_6^4+B_6^6O_6^6\nonumber\\
 &&+g_J \mu_{\rm B} \sum_{i=x,y,z}J_iH_i.
 \end{eqnarray}

Here, $B_m^n$ are the CEF parameters and $O_m^n$ are the Stevens operators.~\cite{Stevens_1952, HUTCHINGS1964}
\begin{eqnarray}
O_2^0&=&3J_z^2-J(J+1)\\
O_2^2&=&\frac{1}{2}(J_+^2+J_-^2)\\
O_4^0&=&35J_z^4-30J(J+1)J_z^2+25J_z^2-6J(J+1)\nonumber\\
&&+3J^2(J+1)^2\\
O_4^2&=&\frac{1}{4}\{[7J_z^2-J(J+1)-5](J_+^2+J_-^2)\nonumber\\
&&+(J_+^2+J_-^2)[7J_z^2-J(J+1)-5](J_+^2+J_-^2)]\}\\
O_4^4&=&\frac{1}{2}(J_+^4+J_-^4)\\
O_6^0&=&231J_z^6-315J(J+1)J_z^4+735J_z^4+105J^2(J+1)^2J_z^2\nonumber\\
&&-525J(J+1)J_z^2+294J_z^2-5J^3(J+1)^3\nonumber\\
&&+40J^2(J+1)^2-60J(J+1)\\
O_6^2&=&\frac{1}{4}\{[33J_z^4-18J_z^2J(J+1)-123J_z^2+J^2(J+1)^2\nonumber\\
&&+10J(J+1)+102](J_+^4+J_-^4)\nonumber\\
&&+(J_+^4+J_-^4)[33J_z^4-18J_z^2J(J+1)\nonumber\\
&&-123J_z^2+J^2(J+1)^2+10J(J+1)+102]\}\\
O_6^4&=&\frac{1}{4}\{[11J_z^2-J(J+1)-38](J_+^4+J_-^4)\nonumber\\
&&+(J_+^4+J_-^4)[11J_z^2-J(J+1)-38]\}\\
O_6^6&=&\frac{1}{2}[(J_x+iJ_y)^6+(J_x-iJ_y)^6]
\end{eqnarray}

The numerical values of $B_m^n$, which were used in the present analysis, are listed in TABLE \ref{tab:SCHEME0}, \ref{tab:SCHEME1}, and \ref{tab:SCHEME2}. The second term of Eq. (1) is explained in terms of electric multipole–strain interaction. Particularly for a rank-2 multipole (quadrupole), this term is written as
\begin{eqnarray}
\mathcal{H}_{\rm MS}&=&-g_{\Gamma 1}\bigl[O_{B}\epsilon_{B}+O_{2}^{0}\epsilon_{u}\bigr]-g_{\Gamma 5}\bigl[O_{yz}\epsilon_{yz}+O_{zx}\epsilon_{zx}\bigr]\nonumber\\
&&-g_{\Gamma 6}\bigl[O_{xy}\epsilon_{xy}+O_{2}^{2}\epsilon_{v}\bigr],
\end{eqnarray}
where $g_{\Gamma}$ are the coupling constants for the rank-2 multipoles. $O_{\Gamma}$ and $\epsilon_{\Gamma}$ are quadrupole operators and symmetrized strains, respectively. They are listed in TABLE \ref{tab:SYMMETRY}, and the quadrupole operators are defined in section E.\\

\begin{table*}
\caption{\label{tab:SYMMETRY} Symmetry, symmetrized strains and rotation, even parity multipoles, and elastic constants in hexagonal notation.}
\begin{ruledtabular}
\begin{tabular}{llcr}
Symmetry (D$_{\rm 6h}$)&Strain and Rotation&	Multipole&Elastic Constant\\\hline
$\Gamma_1$$\oplus$$\Gamma_6$(A$_{\rm 1g}$$\oplus$E$_{\rm 2g}$)&$\epsilon_{xx},\epsilon_{yy}$&-&$C_{11}$\\
$\Gamma_1$(A$_{\rm 1g}$)&$\epsilon_{zz}=\epsilon_{B}/3-\epsilon_{u}/\sqrt{3}$&-&$C_{33}$\\
$\Gamma_5$(E$_{\rm 1g}$)&$\epsilon_{yz}$&$O_{yz}=\sqrt{3}(J_{y}J_{z}+J_{z}J_{y})/2$&$C_{44}$\\
&$\epsilon_{zx}$&$O_{zx}=\sqrt{3}(J_{z}J_{x}+J_{x}J_{z})/2$&$C_{44}$\\
$\Gamma_6$(E$_{\rm 2g}$)&$\epsilon_{xy}$&$O_{xy}=\sqrt{3}(J_{x}J_{y}+J_{y}J_{x})/2$&$C_{66}$\\
&$\epsilon_{v}=\epsilon_{xx}-\epsilon_{yy}$&$O_{v}=\sqrt{3}(J_{x}^2-J_{y}^2)/2$&$C_{66}$\\
\hline
$\Gamma_2$(A$_{\rm 2g}$)&$\omega_{\rm xy}$&$H_z^{\alpha}=\sqrt{35}(J_+^4+J_-^4)/4i$&$C_{66}$\\
\end{tabular}
\end{ruledtabular}

\caption{\label{tab:CEFp}Fit parameters for the elastic constant $C_{66}$ in the present analysis}
\begin{tabular}{@{}lcccccccccr}\toprule
CEF models&$C^0_{66{\rm(PM)}}({\rm J/m^3})$&$s$&&$t$&$\left|g_{\Gamma6{\rm(PM)}}\right|({\rm K})$&$g'_{\Gamma6{\rm(PM)}}({\rm K})$&$C^0_{66{\rm(AFM)}}({\rm J/m^3})$&$\left|g_{\Gamma6{\rm(AFM)}}\right|$&$g'_{\Gamma6{\rm(AFM)}}$\\
\colrule
Scheme 0		&9.1062 &  0.28  &&  180  &30  &-	&-		&-	&-& Oyamada {\it et al.}~\cite{Oyamada2009}\\
Scheme 1 and 2	&9.1062 &  0.28  &&  180  &15.5 &+0.42&9.101&21.0&-0.045& present work
\\
\botrule
\end{tabular}
\end{table*}

\begin{table*}
\caption{\label{tab:SCHEME0}CEF parameters and wave functions for CEF Scheme 0~\cite{Oyamada2009}: Hexagonal ($J = 9/2$)}
\begin{tabular}{@{}lcccccccccccc}\toprule
\colrule
CEF parameters&Scheme 0&&&&$B_2^0$(K)&$B_2^2$(K)&$B_4^0$(K)&$B_4^2$(K)&$B_6^0$(K)&$B_6^2$(K)&$B_6^6$(K)& Ref.\\
\colrule
Hexagonal&$J = 9/2$ &&&&+1.0&-&-0.135&-&+0.02&-&+0.3&\\
\\
Eigenvalues&Wave functions&&&&&&&\\
\colrule
E (K)&Symmetry&&$\left|+9/2\right>$&$\left|+7/2\right>$&$\left|+5/2\right>$&$\left|+3/2\right>$&$\left|+1/2\right>$&$\left|-1/2\right>$&$\left|-3/2\right>$&$\left|-5/2\right>$&$\left|-7/2\right>$&$\left|-9/2\right>$\\
\colrule
0	&	$\Gamma_8^{(1)}$	&&			&$-\kappa$	& 		&		&		&		&		&$\lambda$&		&		\\
0	&	$\Gamma_8^{(1)}$	&&			&		&$\lambda$&		&		&		&		&		&$-\kappa$&		\\
616.2&	$\Gamma_7$&&&&		&		&+1		&		&		&		&		&		\\	
616.2&	$\Gamma_7$&&&&		&		&		&+1		&		&		&		&		\\		
982.6&	$\Gamma_9^{(2)}$	&&$\nu$		&		&		&		&		&		&$\mu$	&		&		&		\\		
982.6&	$\Gamma_9^{(2)}$	&&			&		&		&$\mu$	&		&		&		&		&		&$\nu$	\\		
3006&	$\Gamma_9^{(1)}$	&&$\mu$		&		&		&		&		&		&$-\nu$&		&		&		\\
3006&	$\Gamma_9^{(1)}$	&&			&		&		&		&$\nu$	&		&		&		&		&$\mu$	\\
3649&	$\Gamma_8^{(2)}$	&&			&		&$\kappa$&		&		&		&		&		&$\lambda$&		\\
3649&	$\Gamma_8^{(2)}$	&&			&$\lambda$&		&		&		&		&		&$\kappa$&		&		\\
\botrule
$\kappa=0.8831$\\
$\lambda=0.4692$\\
$\nu=0.7768$\\
$\mu=-0.6298$\\
\end{tabular}
\end{table*} 
\begin{table*}
\caption{\label{tab:SCHEME1}CEF parameters and wave functions for CEF Scheme 1 (Present work): Hexagonal ($J = 4$)}
\begin{tabular}{@{}lccccccccccc}\toprule
\colrule
CEF parameters&Scheme 1&&$B_2^0$(K)&$B_2^2$(K)&$B_4^0$(K)&$B_4^2$(K)&$B_4^4$(K)&$B_6^0$(K)&$B_6^2$(K)&$B_6^4$(K)&$B_6^6$(K)\\
\colrule
Hexagonal&$J = 4$ &&+1.0&-&-0.02&-&-&-0.005&-&-&+0.32\\
\\
Eigenvalues&Wave functions&&&&&&\\
\colrule
E (K)&Symmetry&&$\left|+4\right>$&$\left|+3\right>$&$\left|+2\right>$&$\left|+1\right>$&$\left|0\right>$&$\left|-1\right>$&$\left|-2\right>$&$\left|-3\right>$&$\left|-4\right>$\\
\colrule
0	&	$\Gamma_5^{(2)}$	&&$-\alpha$	&			& 			&		&		&		&$+\beta$&			&		\\
0	&	$\Gamma_5^{(2)}$	&&			&			&$-\beta$	&		&		&		&		&			&$-\alpha$\\
19.1&	$\Gamma_4$		&&			&+1/$\sqrt{2}$&			&		&		&		&		&-1/$\sqrt{2}$&		\\	
652.1&	$\Gamma_6$		&&			&			&			&+1		&		&		&		&			&		\\		
652.1&	$\Gamma_6$		&&			&			&			&		&		&-1		&		&			&		\\
770.6&	$\Gamma_1$		&&			&			&			&		&+1		&		&		&			&		\\		
1225&	$\Gamma_5^{(1)}$	&&$-\beta$	&			&			&		&		&$-\alpha$&		&			&		\\		
1225&	$\Gamma_5^{(1)}$	&&			&			&$+\alpha$	&		&		&		&		&			&$+\beta$\\
1632&	$\Gamma_3$		&&			&+1/$\sqrt{2}$&			&		&		&		&		&+1/$\sqrt{2}$&		\\
\botrule
$\alpha=0.6718$\\
$\beta=0.7408$\\
\end{tabular}
\end{table*}
\begin{table*}
\caption{\label{tab:SCHEME2}CEF parameters and wave functions for CEF Scheme 2 (Present work): Orthorhombic ($J = 4$)}
\begin{tabular}{@{}lccccccccccc}\toprule
\colrule
CEF parameters&Scheme 1&&$B_2^0$(K)&$B_2^2$(K)&$B_4^0$(K)&$B_4^2$(K)&$B_4^4$(K)&$B_6^0$(K)&$B_6^2$(K)&$B_6^4$(K)&$B_6^6$(K)\\
\colrule
Ortho.&$J = 4$ &&+1.0&+0.030&-0.02&+0.001&0&-0.005&0&0&+0.32\\
\\
Eigenvalues&Wave functions&&&&&&\\
\colrule
E (K)&Symmetry&&$\left|+4\right>$&$\left|+3\right>$&$\left|+2\right>$&$\left|+1\right>$&$\left|0\right>$&$\left|-1\right>$&$\left|-2\right>$&$\left|-3\right>$&$\left|-4\right>$\\
\colrule
0	&	$\Gamma_1^{(52)}$	&&$-\alpha'$	&			&$+\beta'$	&			&-2$\delta$	&			&$+\beta'$&			&$-\alpha'$\\
0.79&	$\Gamma_2^{(52)}$	&&$-\alpha'$	&			&$-\beta'$	&			&			&			&$+\beta'$&			&$+\alpha'$\\
19.49&	$\Gamma_4$		&&			&-1/$\sqrt{2}$&			&+4$\delta$	&			&-4$\delta$&		&+1/$\sqrt{2}$&		\\	
652.4&	$\Gamma_2^{6}$		&&			&-4$\delta$	&			&-1/$\sqrt{2}$&			&+1/$\sqrt{2}$&		&+4$\delta$	&		\\		
652.6&	$\Gamma_1^{6}$		&&			&-2$\delta$	&			&+1/$\sqrt{2}$&			&+1/$\sqrt{2}$&		&-2$\delta$	&		\\
771.0&	$\Gamma_1^{(1)}$		&&+3$\delta$	&			&+$\delta$	&			&-1			&			&+$\delta$&			&+3$\delta$\\		
1225&	$\Gamma_1^{(51)}$	&&$+\beta'$	&			&$-\alpha'$	&			&			&			&$+\alpha'$&			&$-\beta'$\\		
1226&	$\Gamma_2^{(51)}$	&&$+\beta'$	&			&$+\alpha'$	&			&+4$\delta$	&			&$+\alpha'$&			&$+\beta'$\\
1632&	$\Gamma_3$		&&			&-1/$\sqrt{2}$&			&-2$\delta$	&			&-2$\delta$	&		&+1/$\sqrt{2}$&		\\
\botrule
$\alpha'=0.4750$\\
$\beta'=0.5238$\\
$\delta=0.0001$\\
\end{tabular}
\end{table*}
Then, the free energy of the local $5f$-electronic states in the CEF can be written as
\begin{eqnarray}
F=U-Nk_{\rm B}T\ln \sum_{i}\exp\{-E_i(\epsilon_\Gamma)/k_{\rm B}T\}.
\end{eqnarray}
Here, $N$ is the number of U ions in a unit volume, $E_n$($\epsilon_\Gamma$) is a perturbed CEF level as a function of strain and $n$ is a collective index for $J$ multiplets and their degenerate states. $U$ gives the internal energy of the strained system in zero magnetic field, which is written in terms of the symmetry strains and elastic constants listed in TABLE \ref{tab:SYMMETRY} as,
\begin{eqnarray}
U&=&\frac{1}{2}\bigl[C_B\epsilon_B^2+C_{Bu}\epsilon_B\epsilon_u+C_u\epsilon_u^2+C_{66}(\epsilon_{xy}^2+\epsilon_v^2)\nonumber\\
&&+C_{44}(\epsilon_{yz}^2+\epsilon_{zx}^2)\bigr].
\end{eqnarray}
Here, $C_{Bu} =-\frac{(C_{11}^0+C_{12}^0-C_{13}^0-C_{14}^0 )}{\sqrt 3}$. The contribution of the electric hexadecapole $H_z^{\alpha}$ with $\Gamma_2$ symmetry is not considered for zero magnetic field, since it will couple to the transverse $C_{66}$ mode only in a finite magnetic field due to the rotational invariance~\cite{Fulde, Goto}. In the framework of second-order perturbation theory, the temperature dependence of the elastic constant is given by
\begin{equation}
C_\Gamma(T,H)=C_\Gamma^0-Ng_\Gamma^2\chi_\Gamma(T, H).
\end{equation}

Here, $C_\Gamma^0$ is the background of the elastic constant. The single-ion quadrupolar susceptibility $\chi_\Gamma$ is defined as the second derivative of the free energy in the limit $\epsilon_\Gamma\rightarrow 0$,
\begin{eqnarray}
-g_\Gamma^2\chi_\Gamma &\equiv & \frac{\partial^2 F}{\partial\epsilon_\Gamma^2}\Bigr|_{\epsilon_\Gamma \rightarrow 0}\nonumber\\
&=&-\left<\frac{\partial^2E_i}{\partial\epsilon_\Gamma^2}\right>
+\frac{1}{k_{\rm B}T}\Bigl[\left<\Bigl(\frac{\partial E_i}{\partial\epsilon_\Gamma}\Bigr)^2 \right>-\left< \frac{\partial E_i}{\partial\epsilon_\Gamma}\right>^2\Bigr].\nonumber\\
\end{eqnarray}
Here, the brackets $\left< \;\; \right>$ indicate the thermal average. In addition to the strain-quadrupole interaction (Eq. 12), the intersite quadrupole-quadrupole interaction can also be added by using the molecular-field approximation of the quadrupolar moment $O_{\Gamma}$ by considering sublattices $\alpha$, as

\begin{eqnarray}
\mathcal{H}_{\rm MM}&=&-\sum_{\alpha}\sum_{\Gamma}g'_{\Gamma}\left<O_{\Gamma}\right>O_{\Gamma}^{\alpha}\nonumber\\
&=&-g'_{\Gamma 1}\left<O_{2}^0\right>O_{2}^{0(\alpha)}-g'_{\Gamma 5}\Bigl\{\left<O_{yz}\right>O_{yz}^{(\alpha)}+\left<O_{zx}\right>O_{zx}^{(\alpha)}\Bigr\}\nonumber\\
&&-g'_{\Gamma 6}\Bigl\{\left<O_{xy}\right>O_{xy}^{(\alpha)}+\left<O_{2}^2\right>O_{2}^{2(\alpha)}\Bigr\},
\end{eqnarray}
\begin{eqnarray}
\mathcal{H}_{\rm MS}+\mathcal{H}_{\rm MM}&=&-\sum_{\alpha}\Bigl\{-g_{\Gamma}\sum_{\Gamma\gamma}O_{\Gamma\gamma}^{(\alpha)}\epsilon_{\Gamma\gamma}+g'_{\Gamma}\sum_{\Gamma\gamma}\left<O_{\Gamma\gamma}\right>O_{\Gamma\gamma}^{\alpha}\Bigr\}\nonumber\\
&=&-g_{\Gamma}\sum_{\alpha}\Bigl\{\sum_{\Gamma\gamma}O_{\Gamma\gamma}^{(\alpha)}\Bigl[\epsilon_{\Gamma\gamma}+\frac{g'_{\Gamma}}{g_{\Gamma}}\left<O_{\Gamma}\right>\Bigr]\Bigr\}\nonumber\\
&=&-g_{\Gamma}\sum_{\alpha}\Bigl\{\sum_{\Gamma\gamma}O_{\Gamma\gamma}^{(\alpha)}\epsilon_{\Gamma\gamma}^{\rm eff}\Bigr\}.
\end{eqnarray}
\begin{equation}
\epsilon_{\Gamma\gamma}^{\rm eff}=\epsilon_{\Gamma\gamma}+\frac{g'_\Gamma}{g_\Gamma}\left<O_{\Gamma}\right>.
\end{equation}
The temperature dependence of the elastic constant (Eq. 15) can be rewritten as,
\begin{equation}
C_\Gamma(T,H)=C_\Gamma^0-\frac{Ng_\Gamma^2\chi_\Gamma(T, H)}{1-g'_\Gamma\chi_\Gamma(T, H)}.
\end{equation}
In the present analysis, the second term is divided into two terms with regard to the different U sites: 2/3 U ions at the U$_{\rm AFM}$ sites [Fig. 1(b) in the main text] and 1/3 of U ions at the U$_{\rm PM1}$ and U$_{\rm PM2}$ sites. We defined two independent sets of coupling constants for each U-ion contribution as

\begin{eqnarray}
C_\Gamma(T,H)&=&C_{66(\rm PM)}-\frac{\frac{2}{3}Ng_{\Gamma6(\rm PM)}^2\chi_{\Gamma6}(T)}{1-g'_{\Gamma6(\rm PM)}\chi_{\Gamma6}(T)}\nonumber\\
&&-\frac{\frac{1}{3}Ng_{\Gamma6(\rm AFM)}^2\chi_{\Gamma6}(T)}{1-g'_{\Gamma6(\rm AFM)}\chi_{\Gamma6}(T)}.
\end{eqnarray}
Here, the phonon background in the PM phase is redefined as the phenomenological fit $C_{66 {\rm(PM)}}(T)=C_{66 {\rm(PM)}}^0-\frac{s}{\exp(t/T)-1}$ with parameters $C_{66 {\rm(PM)}}^0$, $s$, and $t$ (listed in TABLE \ref{tab:CEFp}). $\left|g_{\Gamma6{\rm(PM)}}\right|$ and $\left|g_{\Gamma6{\rm(AFM)}}\right|$ are the coupling constants of the quadrupolar-strain interaction for the PM and AFM phases, respectively. The coupling constant of the quadrupolar intersite interaction $g'_{\Gamma6(\rm(PM)}$ and $g'_{\Gamma6(\rm AFM)}$ are defined as the same manner. For the AFM phase, we assume that the second term of Eq. (21), the quadrupolar contribution from U$_{\rm(AFM)}$ is a temperature-independent constant value, and redefine the temperature-independent background as $C^0_{66 {\rm(AFM)}}$.
\begin{equation}
C_\Gamma(T,H)=C_{66(\rm AFM)}^0-\frac{\frac{1}{3}Ng_{\Gamma6(\rm AFM)}^2\chi_{\Gamma6}(T)}{1-g'_{\Gamma6(\rm AFM)}\chi_{\Gamma6}(T)}.
\end{equation}

\section*{\label{sec:Definition}E.	Definition of Multipolar Moments and Equivalent Operator Expressions}
The electric multipolar operators are defined by multipolar expansion of the electrostatic potential~\cite{Kusunose2008} as
\begin{equation}
Q_{lm}\equiv e\sum_{j=1}^{n_f}r_j^i Z_{lm}^*(r_j).
\end{equation}
Here, $e < 0$ is the electron charge, $n_f$ is the number of $f$ electrons. $Z_{lm}(r_j)$ is written by using spherical harmonics $Y_{lm}(r_j)$ as
\begin{equation}
 Z_{lm}(r_j)\equiv \sqrt{4\pi/(2l+1)}Y_{lm}(r_j).
\end{equation}
Eq. (24) can be rewritten by replacing ($x, y, z$) in $Z_{lm}$ with spherical tensor operators $J_{lm}$ with following transformations,
\begin{equation}
x^{n_x}y^{n_y}z^{n_z}\rightarrow \frac{n_x!n_y!n_z!}{(n_x+n_y+n_z)!}\sum_{\mathcal{P}}\mathcal{P}(J_x^{n_x}J_y^{n_y}J_z^{n_z}).
\end{equation}
Here, $\mathcal{P}$ is a sum of all possible permutations. The operator $J_{lm}$ has the following commutation relations, with the ladder operator $J_{\pm}=J_x \pm iJ_y$:
\begin{equation}
J_{ll}=(-1)^l\sqrt{\frac{(2l-1)!!}{(2l)!!}}(J_+)^l,
\end{equation}
\begin{equation}
[J_{-},J_{lm}]=\sqrt{(l+m)(l-m+1)}J_{lm-1}.
\end{equation}
Following are quadrupolar and hexadecapolar operators, which are used in the present analysis:\\
{\it i) Rank 2 (Quadrupole)}
\begin{eqnarray}
\Gamma_5(\rm E_{1g}) :&\nonumber\\
O_{yz}=&\frac{i}{\sqrt{2}}[J_{21}+J_{2-1}]=\frac{\sqrt{3}}{2}(J_yJ_z+J_zJ_y)
\end{eqnarray}
\begin{eqnarray}
\Gamma_5(\rm E_{1g}) :&\nonumber\\
O_{zx}=&\frac{1}{\sqrt{2}}[-J_{21}+J_{2-1}]=\frac{\sqrt{3}}{2}(J_zJ_x+J_xJ_z)
\end{eqnarray}\\
\begin{eqnarray}
\Gamma_6(\rm E_{\rm 2g}) :&\nonumber\\
O_{xy}=&\frac{i}{\sqrt{2}}[-J_{22}+J_{2-2}]=\frac{\sqrt{3}}{2}(J_xJ_y-J_yJ_x)
\end{eqnarray}
\begin{eqnarray}
\Gamma_6(\rm E_{\rm 2g}) :&\nonumber\\
O_v=&\frac{\sqrt{3}}{2}O_2^2=\frac{i}{\sqrt{2}}[J_{22}+J_{2-2}]=\frac{\sqrt{3}}{2}(J_x^2+J_y^2)
\end{eqnarray}
{\it ii) Rank 4 (Hexadecapole)}\\
\begin{eqnarray}
\Gamma_2(\rm A_{\rm 2g}) :&&\nonumber\\
H_z^{\alpha}&=&\frac{\sqrt{35}}{4i}[-J_{44}+J_{4-4}]\nonumber\\
&=&\frac{\sqrt{35}}{8}\{(J_x^3J_y+J_x^2J_yJ_x+J_xJ_yJ_x^2+J_yJ_x^3)\nonumber\\
&-&(J_xJ_y^3+J_y^2J_xJ_y+J_yJ_xJ_y^2+J_xJ_y^3)\}
\end{eqnarray}\\

\section*{\label{sec:Matrix}F.	Matrix Elements of Magnetic dipole and electric quadrupoles }
Here, we show the matrix elements of magnetic dipole $J_x$ and $J_z$, and electric quadrupoles $O_{xy}$ and $O_{yz}$, which are calculated by using present CEF models. Especially, the finite diagonal- and off-diagonal-matrix elements of $J_x$ and $O_{xy}$ are correspond to the transitions indicated by arrows in Fig. 3(a) of the main text.\\

i) CEF Scheme 0 ($J = 9/2$)\\
\\
\begin{equation}
\begin{array}{cll}
 & {\rm Energy}&{\rm \hspace{45pt} Wavefunction}\\
\hline
 &0\,{\rm K} & \left|\Gamma_8^{(1)}\pm\right> = -\kappa\left|\pm7/2\right>+\lambda\left|\mp5/2\right>\\
 &616.2\,{\rm K}& \left|\Gamma_7\pm\right> = \left|\pm 1\right>\\
 & &\\
 & &\kappa=0.8831\\
 & &\lambda=0.4692\\
 & &\nu=0.7768\\
 & &\mu=-0.6298\\
\end{array}
\end{equation}
\begin{equation}
\left<J_x \right> =\left[\begin{array}{rcc|cc}
&\left|\Gamma_8^{(1)}+\right> &\left|\Gamma_8^{(1)}-\right> &\left|\Gamma_7+\right> &\left|\Gamma_7-\right>\\
\left<\Gamma_8^{(1)}+\right|	&0 		& -1.675&0 		& 0		\\
\left<\Gamma_8^{(1)}-\right|	&-1.675 & 0 		&0 		& 0		\\
\hline
\left<\Gamma_7+\right|	&0 		& 0 		&-0.5 	& 0		\\
\left<\Gamma_7-\right|	&0 		& 0 		&0 		& +0.5	\\

\end{array}\right]\nonumber
\end{equation}

\begin{equation}
\left<J_z \right> =\left[\begin{array}{rcc|cc}
&\left|\Gamma_8^{(1)}+\right> &\left|\Gamma_8^{(1)}-\right> &\left|\Gamma_7+\right> &\left|\Gamma_7-\right>\\
\left<\Gamma_8^{(1)}+\right|&-2.179 & 0 &0 & 0\\
\left<\Gamma_8^{(1)}-\right|&0 & +2.179 &0 & 0\\
\hline
\left<\Gamma_7+\right|&0 & 0 &0 & +2.5\\
\left<\Gamma_7-\right|	&0 & 0 &+2.5 & 0\\
\end{array}\right]\nonumber
\end{equation}

\begin{equation}
\left<O_{xy} \right> =\left[\begin{array}{rcc|cc}
&\left|\Gamma_8^{(1)}+\right> &\left|\Gamma_8^{(1)}-\right> &\left|\Gamma_7+\right> &\left|\Gamma_7-\right>\\
\left<\Gamma_8^{(1)}+\right|	&0 		& 0		&0 		& -5.267\\
\left<\Gamma_8^{(1)}-\right|	&0		& 0 		&+5.267 	& 0		\\
\hline
\left<\Gamma_7+\right|	&0 		&-5.267 	&0 		& 0		\\
\left<\Gamma_7-\right|	&+5.267 	& 0 	&0 		& 0		\\
\end{array}\right]\times i\nonumber
\end{equation}

\begin{equation}
\left<O_{yz} \right> =\left[\begin{array}{rcc|cc}
&\left|\Gamma_8^{(1)}+\right> &\left|\Gamma_8^{(1)}-\right> &\left|\Gamma_7+\right> &\left|\Gamma_7-\right>\\
\left<\Gamma_8^{(1)}+\right|	&0 		& 0		&0 		& 0		\\
\left<\Gamma_8^{(1)}-\right|	&0		& 0 		&0 		& 0		\\
\hline
\left<\Gamma_7+\right|	&0 		& 0 		&0 		& 0		\\
\left<\Gamma_7-\right|	&0 		& 0 		&0 		& 0	\\
\end{array}\right]\times i\nonumber
\end{equation}
\\
ii) CEF Scheme 1 ($J = 4$)\\

For CEF Scheme 1 and 2, the matrix element of the $\Gamma_6$ doublet at around 652 K is omitted because its effect on the low-temperature properties is negligible.\\
\\
\begin{equation}
\begin{array}{cll}
& {\rm Energy}	& {\rm \hspace{45pt} Wavefunction}\\
\hline
&0\,{\rm K} 	& \left|\Gamma_5^{(2)}\pm\right> = -\alpha\left|\pm4\right>\pm\beta\left|\mp2\right>\\
&19.1\,{\rm K} 	& \left|\Gamma_4\right> = \frac{1}{\sqrt{2}}(\left|+3\right>-\left|-3\right>)\\
&&\\
&&\alpha=0.6718\\
&&\beta=0.7408
\end{array}
\end{equation}
\begin{equation}
\left<J_x \right> =\left[\begin{array}{rcc|c}
&\left|\Gamma_5^{(2)}+\right> &\left|\Gamma_5^{(2)}-\right> &\left|\Gamma_4\right>\\
\left<\Gamma_5^{(2)}+\right|	&0 			& 0			&-1.652 		\\
\left<\Gamma_5^{(2)}-\right|	&0 			& 0 			&-1.652 		\\
\hline
\left<\Gamma_4\right|			&-1.652 		& -1.652 	&0			\\
\end{array}\right]\nonumber
\end{equation}

\begin{equation}
\left<J_z \right> =\left[\begin{array}{rcc|c}
&\left|\Gamma_5^{(2)}+\right> &\left|\Gamma_5^{(2)}-\right> &\left|\Gamma_4\right>\\
\left<\Gamma_5^{(2)}+\right|	&-0.708 		& 0			&0 			\\
\left<\Gamma_5^{(2)}-\right|	&0 			& +0.708 	&0 			\\
\hline
\left<\Gamma_4\right|			&0 			& 0 			&0			\\
\end{array}\right]\nonumber
\end{equation}

\begin{equation}
\left<O_{xy} \right> =\left[\begin{array}{rcc|c}
&\left|\Gamma_5^{(2)}+\right> &\left|\Gamma_5^{(2)}-\right> &\left|\Gamma_4\right>\\
\left<\Gamma_5^{(2)}+\right|	&0 			& +5.266		&0 			\\
\left<\Gamma_5^{(2)}-\right|	&-5.266	 	& 0 			&0 			\\
\hline
\left<\Gamma_4\right|			&0 			& 0 			&0			\\
\end{array}\right]\times i\nonumber
\end{equation}

\begin{equation}
\left<O_{yz} \right> =\left[\begin{array}{rcc|c}
&\left|\Gamma_5^{(2)}+\right> &\left|\Gamma_5^{(2)}-\right> &\left|\Gamma_4\right>\\
\left<\Gamma_5^{(2)}+\right|	&0 			& 0			&-0.198 		\\
\left<\Gamma_5^{(2)}-\right|	&0	 		& 0 			&-0.198 		\\
\hline
\left<\Gamma_4\right|			&+0.198 		& +0.198 	&0			\\
\end{array}\right]\times i\nonumber
\end{equation}
\\
iii) CEF Scheme 2 ($J = 4$)\\
\\
\begin{equation}
\begin{array}{clll}
& {\rm Energy}	& {\rm \hspace{45pt} Wavefunction}\\
\hline
&0\,{\rm K} 		&\left|\Gamma_1^{(52)}\right>=-\alpha'(\left|+4\right>+\left|-4\right>)\\
&					&\hspace{45pt}+\beta'(\left|+2\right>+\left|-2\right>)-2\delta\left|0\right>\\ 
&0.79\,{\rm K} 		&\left|\Gamma_2^{(52)}\right>=-\alpha'(\left|+4\right>-\left|-4\right>)\\
&					&\hspace{45pt}-\beta'(\left|+2\right>-\left|-2\right>)\\ 		
&19.49\,{\rm K} 	& \left|\Gamma_4\right>=-\frac{1}{\sqrt{2}}(\left|+3\right>-\left|-3\right>+4\delta(\left|+1\right>-\left|-1\right>)\\
&&\\
&&\alpha'=0.4750\\
&&\beta'=0.5238\\
&&\delta=0.0001
\end{array}
\end{equation}

\begin{equation}
\left<J_x \right> =\left[\begin{array}{rcc|c}
&\left|\Gamma_1^{(52)}\right> &\left|\Gamma_2^{(52)}\right> &\left|\Gamma_4\right>\\
\left<\Gamma_1^{(52)}\right|	&0 			& 0			&0	 		\\
\left<\Gamma_1^{(52)}\right|	&0 			& 0 			&+2.351	 	\\
\hline
\left<\Gamma_4\right|			&0 			& +2.351		&0			\\
\end{array}\right]\nonumber
\end{equation}

\begin{equation}
\left<J_z \right> =\left[\begin{array}{rcc|c}
&\left|\Gamma_1^{(52)}\right> &\left|\Gamma_2^{(52)}\right> &\left|\Gamma_4\right>\\
\left<\Gamma_1^{(52)}\right|	&0 			& -0.708		&0	 		\\
\left<\Gamma_1^{(52)}\right|	&-0.708	 	& 0 			&0	 		\\
\hline
\left<\Gamma_4\right|			&0 			& 0		 	&0			\\
\end{array}\right]\nonumber
\end{equation}

\begin{equation}
\left<O_{xy} \right> =\left[\begin{array}{rcc|c}
&\left|\Gamma_1^{(52)}\right> &\left|\Gamma_2^{(52)}\right> &\left|\Gamma_4\right>\\
\left<\Gamma_1^{(52)}\right|	&0 			& +5.264		&0	 		\\
\left<\Gamma_1^{(52)}\right|	&-5.264 		& 0 			&0	 		\\
\hline
\left<\Gamma_4\right|			&0 			& 0		 	&0			\\
\end{array}\right]\times i\nonumber
\end{equation}

\begin{equation}
\left<O_{yz} \right> =\left[\begin{array}{rcc|c}
&\left|\Gamma_1^{(52)}\right> &\left|\Gamma_2^{(52)}\right> &\left|\Gamma_4\right>\\
\left<\Gamma_1^{(52)}\right|	&0 			& 0			&0	 		\\
\left<\Gamma_1^{(52)}\right|	&0 			& 0 			&+0.282		\\
\hline
\left<\Gamma_4\right|			&0 			& -0.282		&0			\\
\end{array}\right]\times i\nonumber
\end{equation}

\begin{figure*}
\includegraphics[width=0.8\linewidth]{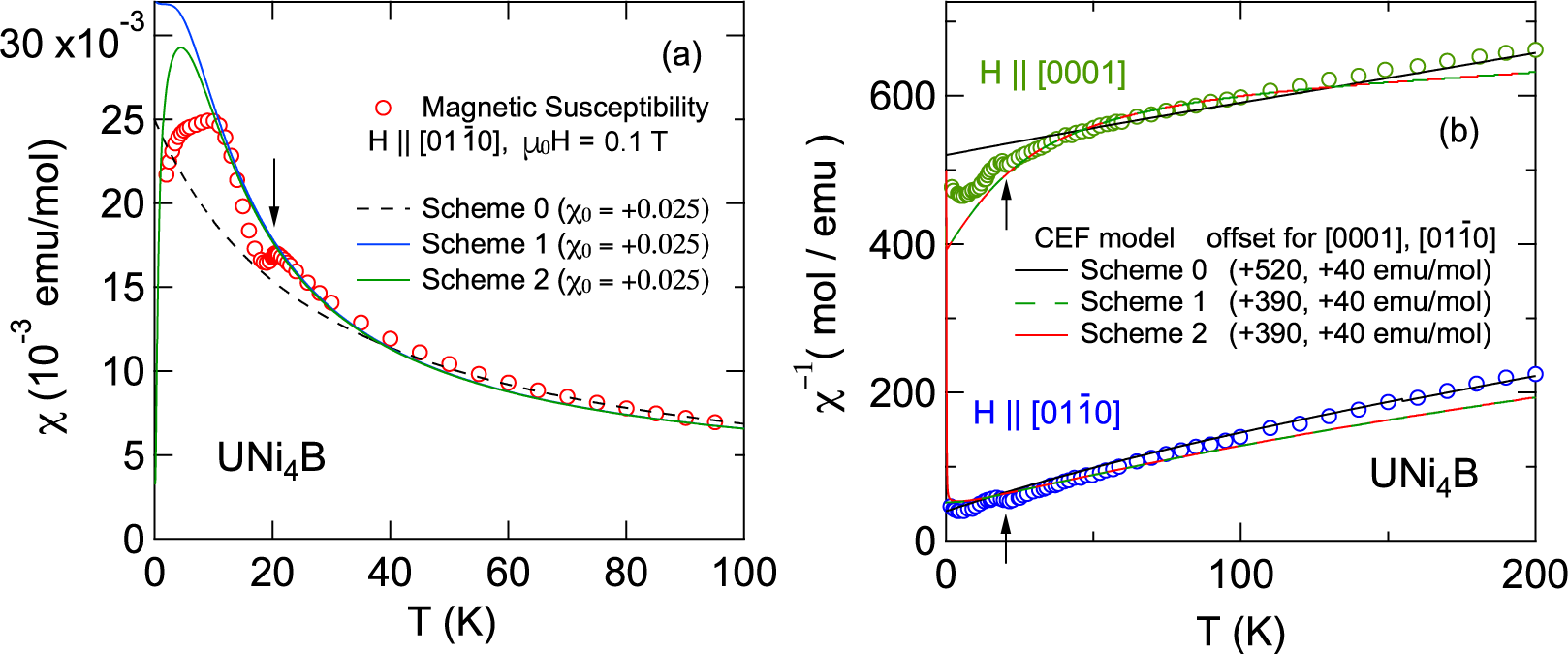}
\caption{\label{fig:figS3}
(a) Measured magnetic susceptibility of UNi$_4$B for $H \parallel[01\bar{1}0]$ (open circles)~\cite{Saito2018} and calculated magnetic susceptibility using the CEF level Schemes 0, 1, and 2. (b) Inverse magnetic susceptibility of UNi$_4$B for $H \parallel[0001]$ and $H \parallel[01\bar{1}0]$  with experimental (open circles) and calculated data.}
\end{figure*}

\section*{\label{sec:Consistency}G.	Consistency Test for CEF Calculation of the Magnetic Susceptibility}
In the main text, we propose new CEF level schemes (Scheme 1 and Scheme 2) to reproduce the elastic response of the $C_{66}$ mode in UNi$_4$B. Here, we show the consistency test using these CEF models by comparing the calculated with the experimentally obtained magnetic susceptibility. Figs. S3(a) and S3(b) show the calculated magnetic susceptibility of $H \parallel[01\bar{1}0]$, and the inverse magnetic susceptibilities for $H \parallel[01\bar{1}0]$ and [0001] compared with the experimental data. We notice that Scheme 0 reproduces the data for the high-temperature region $T > 50$ K well, while Schemes 1 and 2 deviate slightly from the experimental data. On the other hand, Schemes 1 and 2 reproduce the low-temperature region below $\sim50$ K well, including the AFM-ordered phase, and mimic the leveling off and decrease of the data below 10 K for $H \parallel[01{\bar{1}}0]$, while Scheme 0 suggests a continuous increase toward absolute zero. The leveling off and decreasing susceptibility imply the presence of PM U ions in the AFM phase, so the calculated results of Schemes 1 and 2 are consistent with the experimentally obtained magnetic response. This consistency suggests that the change in the Fermi surface due to the AFM order enhances the localized character of the U ions in the AFM phase, and this localized electronic state is reproduced well by the $J = 4$ state. Otherwise, we need to consider a novel coupling mechanism between the strain field and the toroidal magnetic moment (augmented multipole) that originates from parity mixing.\\

\section*{\label{sec:Elastic}H.	Elastic Constant $C_{66}$ and Calculated Quadrupolar Susceptibility at Low Temperatures ($< 1$ K)}
Here, we show additional low-temperatures results ($< 1$ K) obtained using a dilution refrigerator. Figs. S4 and S5 show the relative change of the elastic constant $C_{66}$ as a function of magnetic field and temperature, respectively. The data in Fig. S4 were taken for both increasing and decreasing magnetic field, as indicated by arrows, and the data are vertically shifted for better visualization. The data in Fig. S5 are shifted vertically based on the magnetic-field dependence of $\Delta C_{66}$ to compare the response between $H \parallel[01\bar{1}0]$ and $H \parallel[2\bar{1}\bar{1}0]$. The right- and left-headed horizontal arrows on the side of the magnetic field values indicate heating and cooling processes, respectively. The arrowhead at around 0.3 K indicates the inflection point, defined by the crossing of the linear extrapolations from the low- and high-temperature side. The shallow minimum that appears around 0.3 K in the temperature dependence and around 4 T in the magnetic-field dependence for $H \parallel[2\bar{1}\bar{1}0]$ are twice as large as similar anomalies observed for $H \parallel[01\bar{1}0]$. The anisotropy is roughly consistent with the calculation based on the CEF Scheme 2. Figure S6 shows the calculated $\Gamma_6$(E$_{\rm 2g}$)-quadrupolar susceptibility as a function of temperature at various magnetic fields and as a function of magnetic field at 0.1 K for (a) $H \parallel[01\bar{1}0]$ and (b) $H \parallel[2\bar{1}\bar{1}0]$. Here, Figs. S5 and S6 are displayed in a similar style with corresponding colors. It appears that the local minimum is caused by a level crossing in the magnetic-field dependence of the quadrupolar susceptibility for $H \parallel[2\bar{1}\bar{1}0]$ but not for $H \parallel[01\bar{1}0]$. This anisotropic response of the CEF effect is roughly consistent with the tendency of the local minimum in $C_{66}$, as shown in Fig. S4, where the magnitude of the local minimum for $H \parallel[2\bar{1}\bar{1}0]$ is twice as large as that for $H \parallel[01\bar{1}0]$. In order to reproduce the magnetic-field dependence and anisotropy of $T^*$ to higher fields, a more detailed calculation considering molecular-field models is needed, which is out of the scope of this work. 
\\
\\
\begin{figure*}
\includegraphics[width=0.85\linewidth]{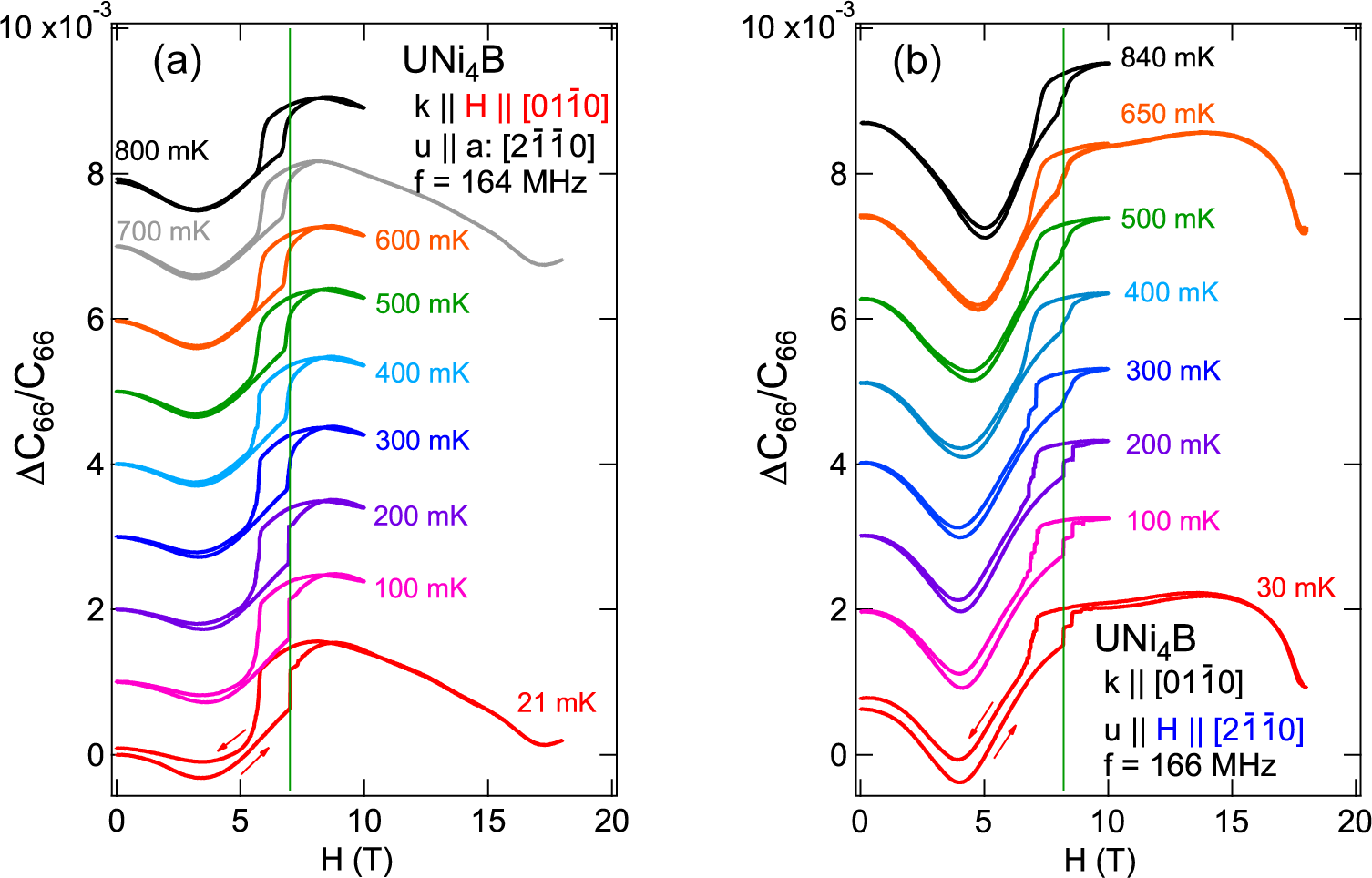}
\caption{\label{fig:figS4}
Relative change of the elastic constant $C_{66}$ of UNi$_4$B as a function of magnetic field for (a) $H \parallel[01\bar{1}0]$ and (b) $H \parallel[2\bar{1}\bar{1}0]$ at various fixed temperatures. }
\end{figure*}
\begin{figure*}
\includegraphics[width=0.85\linewidth]{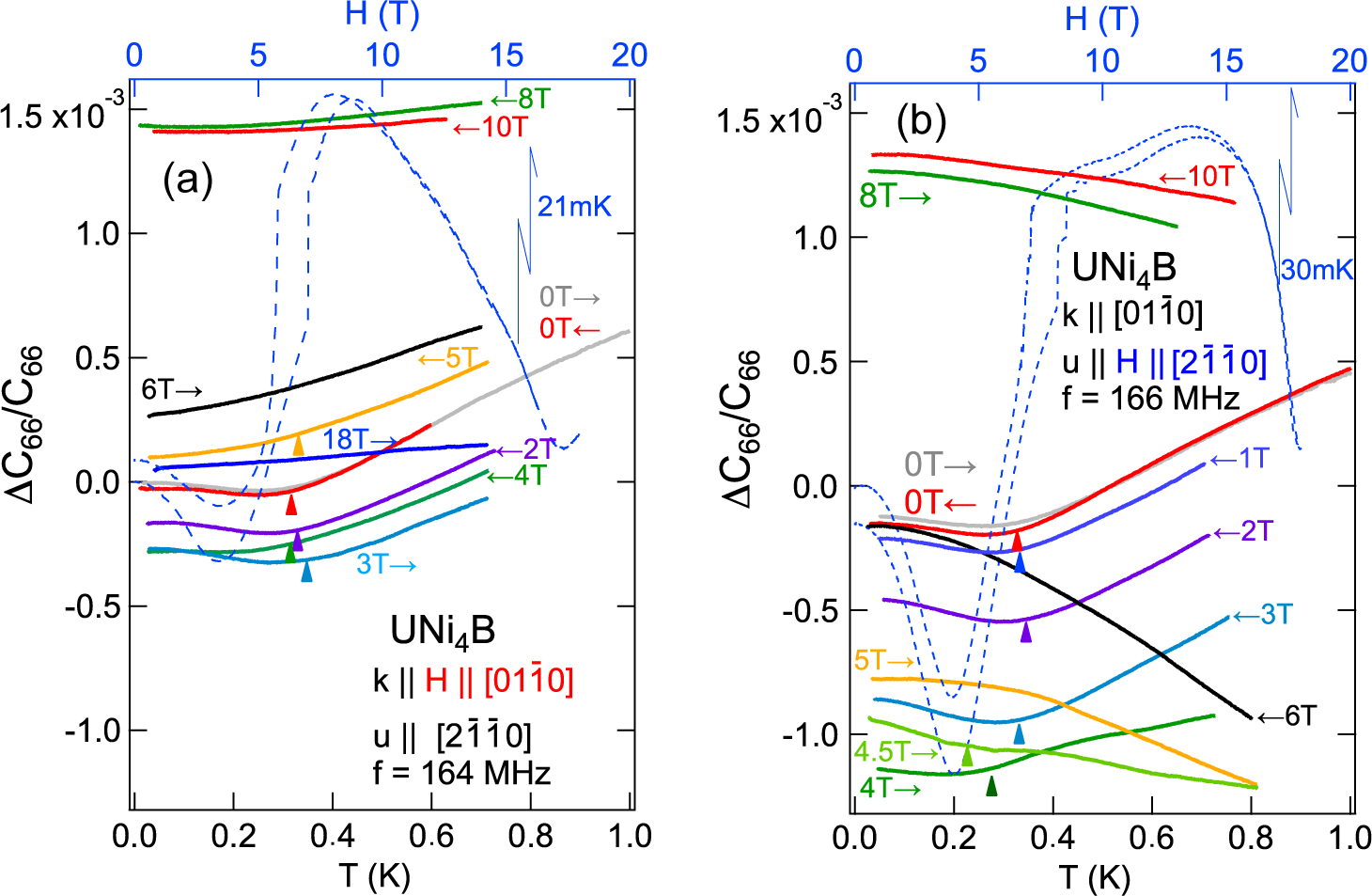}
\caption{\label{fig:figS5}
Relative change of the elastic constant $C_{66}$ of UNi$_4$B as a function of temperature for (a) $H \parallel[01\bar{1}0]$ and (b) $H \parallel[2\bar{1}\bar{1}0]$ at various magnetic fields.}
\end{figure*}
\begin{figure*}
\includegraphics[width=0.85\linewidth]{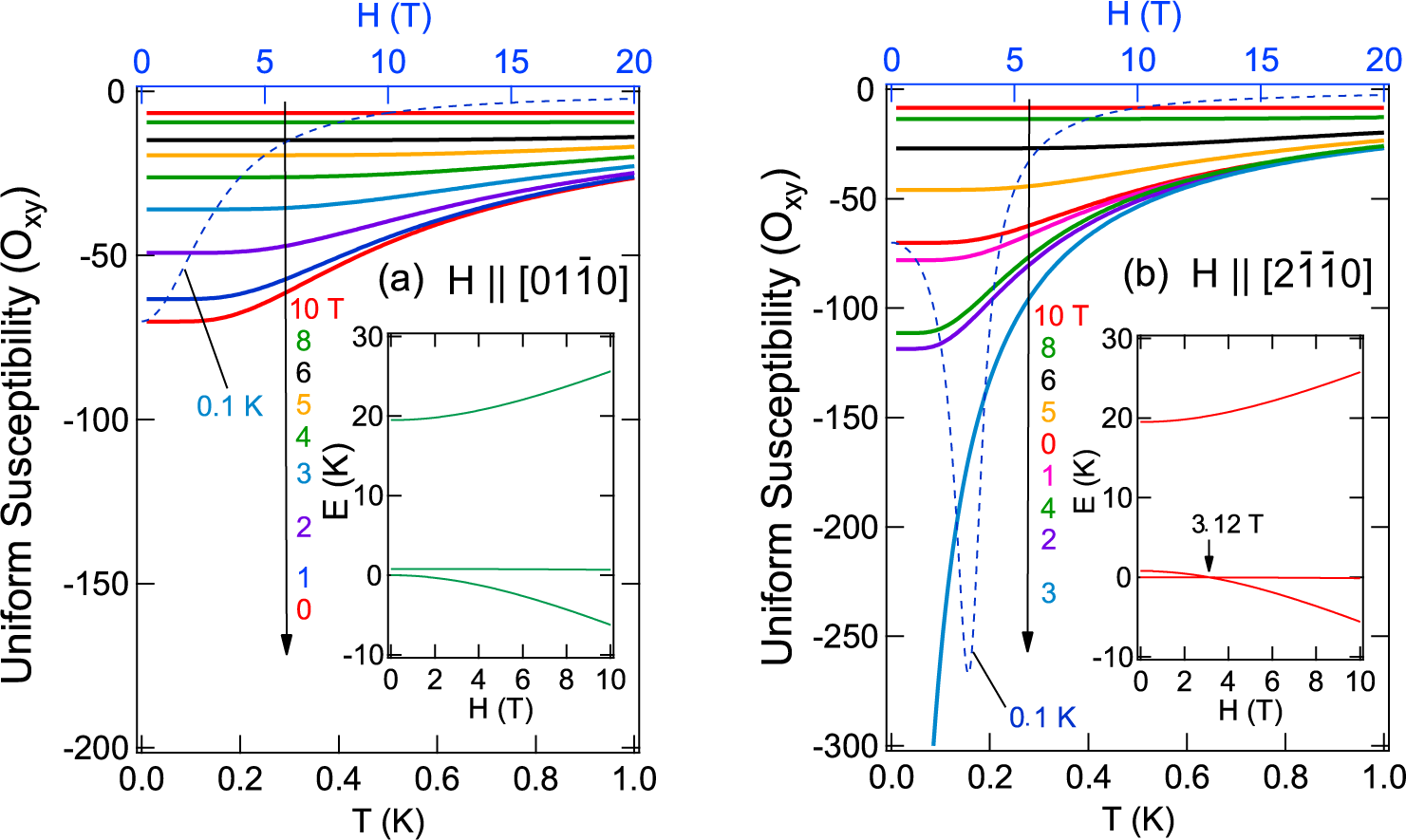}
\caption{\label{fig:figS6}
Calculated $\Gamma_6$ (E$_{\rm 2g}$)-quadrupolar susceptibility of $O_{xy}$ using the CEF Scheme 2 as a function of temperature (bottom axis) and magnetic field (top axis) at 0.1 K for (a) $H \parallel[01\bar{1}0]$ and (b) $H \parallel[2\bar{1}\bar{1}0]$ at various fixed magnetic fields. The inset in each figure shows the magnetic-field dependence of the energy niveaus of the ground-state doublet and the first excited singlet.}
\end{figure*}

\section*{\label{sec:Anisotropy}I. Anisotropy of the $H$-$T$ Phase Diagram and the 330 mK Anomaly}
In Figs. S7 and S8, we display the $H$–$T$ phase diagrams of the low-temperature region on a log and linear $T$ scale for in-plane magnetic fields. The inflection points (Fig. S4), local minima, and elastic anomalies are compiled into the phase diagrams. Open blue symbols represent the phase boundaries for $H \parallel[2\bar{1}\bar{1}0]$, closed red symbols the phase boundaries for $H \parallel[01\bar{1}0]$. The marker (+) with dashed lines indicates the broad minimum in the magnetic-field dependence of $C_{66}$. Figure S8 shows the $H$–$T$ phase diagram for the low-temperature and low-magnetic field region on an enlarged linear-$T$ scale to emphasize the magnetic-field dependence of the $T^*\sim0.33$ K anomaly. The closed red circles ($\bullet$) and open blue squares ($\Box$) represent the temperature where the leveling off or broad minimum appears in $C_{66}$ vs. $T$ in Fig. S5. Open black symbols ($\circ$ and $\Box$) filled with ($+$) or ($\times$) are positions of the specific-heat anomalies previously reported by Movshovich {\it et al.}~\cite{Movshovich1999}, which are displayed for comparison.\\

\begin{figure*}
\includegraphics[width=0.85\linewidth]{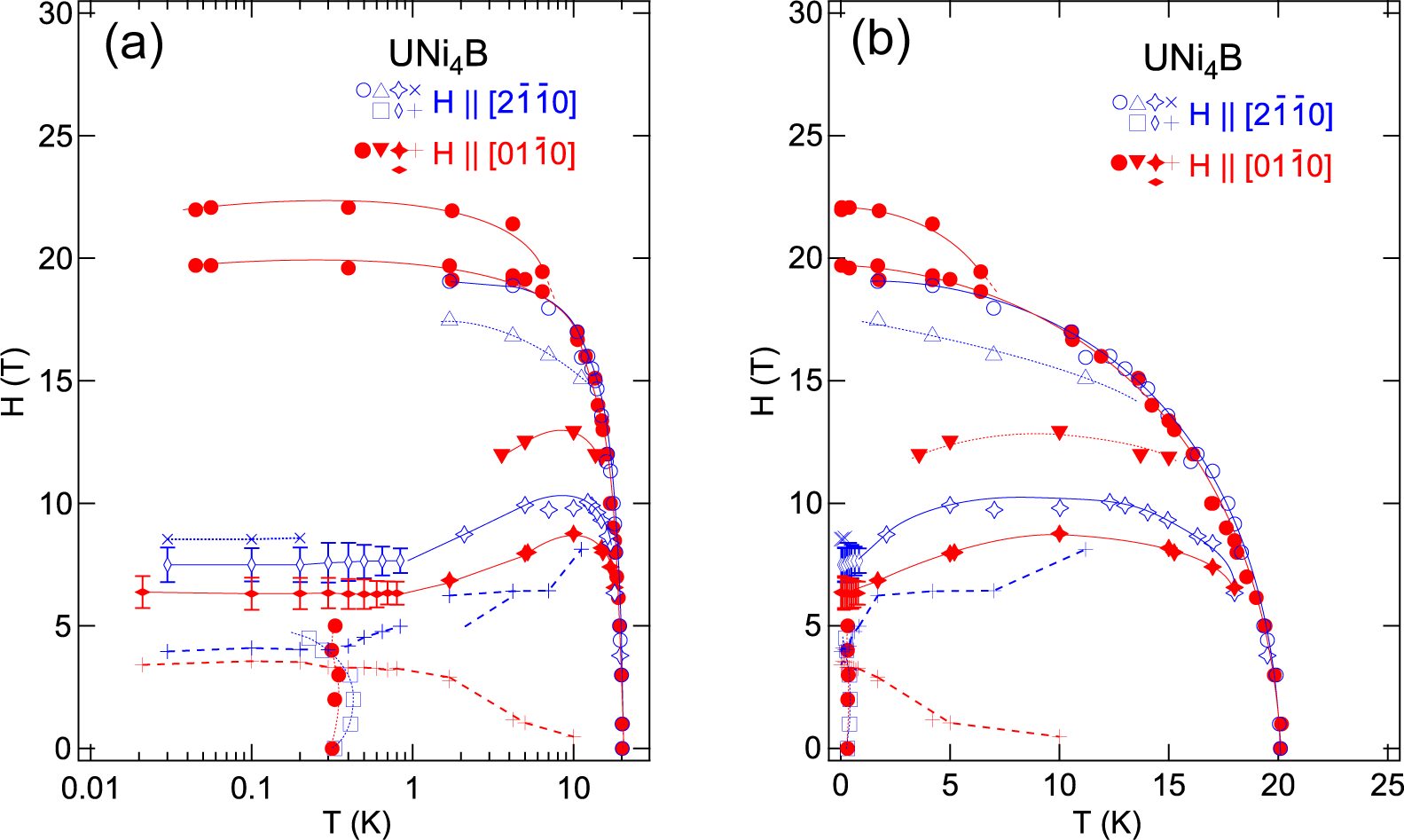}
\caption{\label{fig:figS7}
Magnetic field–temperature phase diagrams of UNi$_4$B for fields aligned along $[2\bar{1}\bar{1}0]$ and $[01\bar{1}0]$ on a (a) log-$T$ and (b) linear-$T$ scale.
} 
\end{figure*}
\begin{figure*}
\includegraphics[width=0.45\linewidth]{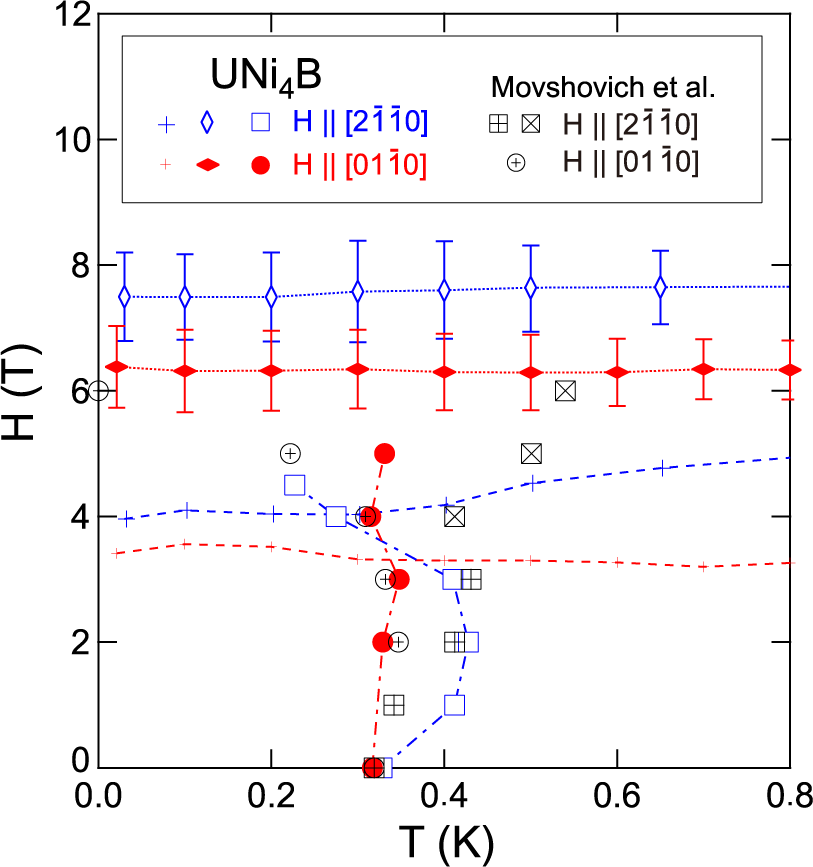}
\caption{\label{fig:figS8}
Magnetic field–temperature phase diagram of UNi$_4$B for the low-temperature and low-magnetic-field region on a linear-$T$ scale.
}
\end{figure*}

\section{\label{sec:Pulse}J. Elastic Response of the $C_{66}$ Mode under Pulsed Magnetic Fields above 30 T}
In Fig. S9, we display the relative change of $C_{66}$ obtained in pulsed magnetic fields. Since the data between 30–60 T shows no further anomalies, we only present the static-field results in the main text. Here, the gray dashed curves marked by, 10 K*, 7 K*, 5 K*, and 1.7 K* show the data obtained in static magnetic fields up to 17 T for comparison.
\\
\begin{figure*}
\includegraphics[width=0.85\linewidth]{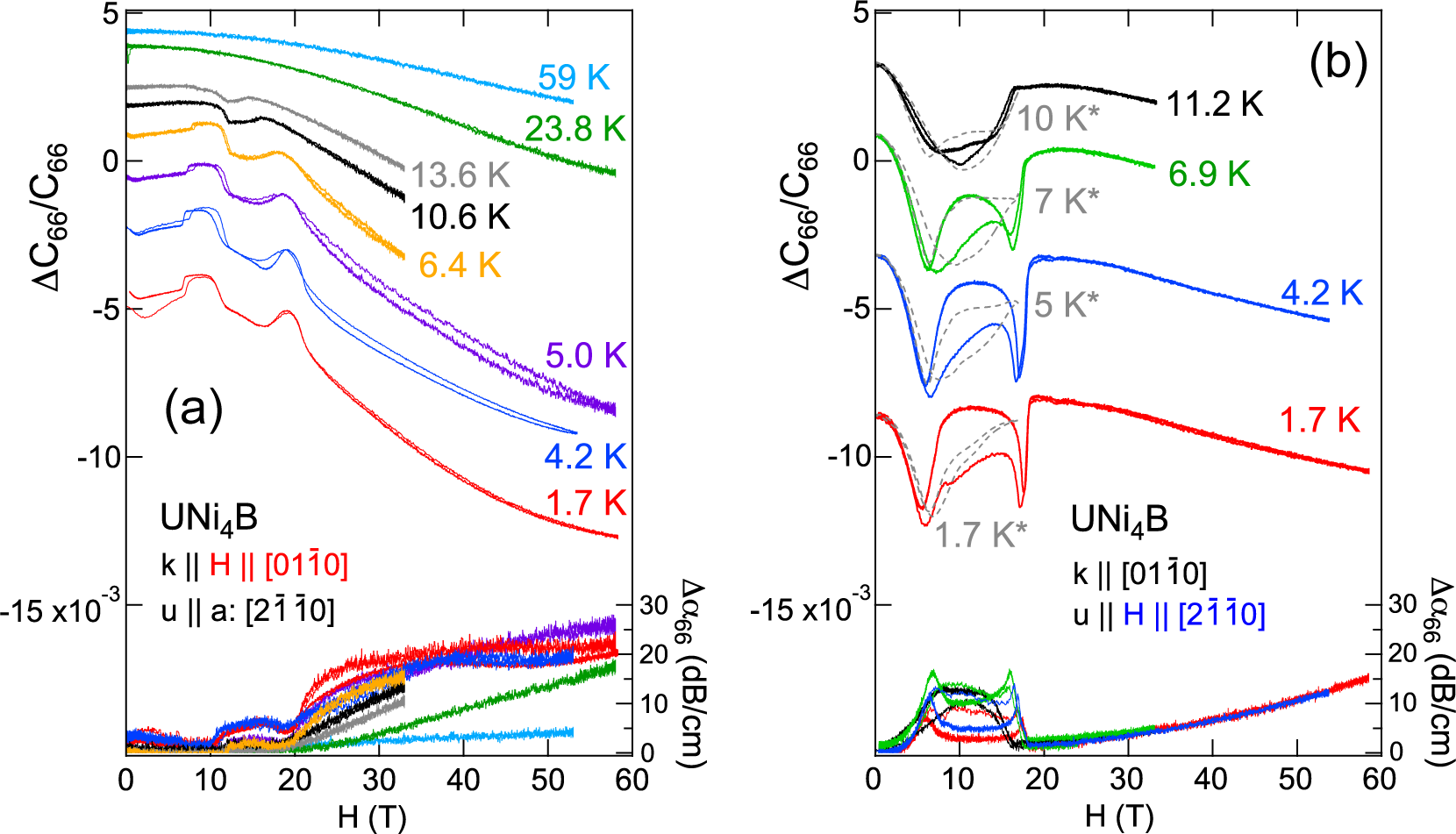}
\caption{\label{fig:figS9}
Magnetic-field dependence of the relative change of $C_{66}$ in pulsed magnetic fields for (a) $H \parallel[01\bar{1}0]$ and (b) $H \parallel[2\bar{1}\bar{1}0]$ at various temperatures. The gray dashed curves marked by 10 K*, 7 K*, 5 K* and 1.7 K* show the data obtained in static magnetic fields up to 17 T for comparison. The lower panels show the sound-attenuation change $\Delta\alpha_{66}$ vs. $H$. All data were taken with both up- and down-field sweeps.
}
\end{figure*}
\end{document}